\documentclass[letterpaper,11pt]{article}
\pdfoutput=1

\usepackage{jheppub} 
\usepackage[dvipsnames]{xcolor}
\definecolor{MMA1}{rgb}{0.368417, 0.506779, 0.709798}
\definecolor{MMA2}{rgb}{0.880722, 0.611041, 0.142051}
\definecolor{MMA3}{rgb}{0.560181, 0.691569, 0.194885}
\definecolor{MMA4}{rgb}{0.922526, 0.385626, 0.209179}
\definecolor{MMA5}{rgb}{0.528488, 0.470624, 0.701351}

\usepackage{subfig}
\usepackage{afterpage}
\usepackage{multirow}
\usepackage{tikz}
\newcommand{\swatch}[1]{\tikz[baseline=-0.6ex] \node[fill=#1,shape=rectangle,draw=#1!10!black,minimum width=5mm
](){};}
\newcommand{\coline}[1]{\tikz{ \draw[#1, ultra thick] (1,0) --(2,0);}}
\newcommand{\dash}[1]{\tikz{ \draw[#1, dashed, ultra thick] (1,0) --(2,0);}}
\newcommand{\dashLong}[1]{\tikz{\draw[#1, dash pattern=on 8pt off 8pt, ultra thick] (1,0) --(2,0);}}
\newcommand{\dashlong}[1]{\tikz{\draw[#1, dash pattern=on 4pt off 4pt, ultra thick] (1,0) --(2,0);}}

\newcommand{\nn}{\nonumber}

\newcommand{\be}{\begin{equation}}
\newcommand{\ee}{\end{equation}}
\newcommand{\bea}{\begin{eqnarray}}
\newcommand{\eea}{\end{eqnarray}}
\newcommand{\system}[1]{\left\{\begin{array}{l} #1\end{array}\right.}

\def\beq#1\eeq{\begin{align}#1\end{align}}
\def\beqnn#1\eeq{\begin{align*}#1\end{align*}}
\newcommand{\rpv}{$R$-parity violation}
\newcommand{\rpa}{$R$-parity}
\newcommand{\rpcing}{$R$-parity conserving\ }
\newcommand{\rpving}{$R$-parity violating\ }

\def \mt {\tilde m}
\def \l {\lambda''}
\newcommand{\fru}[2]{\left( \frac{#1}{\text{#2}}\right)}
\newcommand{\frd}[2]{\left( \frac{\text#1}{{#2}}\right)}
\newcommand{\frud}[2]{\left( \frac{#1}{{#2}}\right)}

\def \tq {{\tilde q}}
\def \tg {\tilde g}

\def \mtt { m_{\tilde t}}

\def \mtq { m_{\tilde q}}
\def \mtg { m_{\tilde g}}
\def \mtc { m_{\tilde \chi}}
\def \tev {\text{~TeV}}
\def \gev {\text{~GeV}}

\def\lijk{\l_{ijk}}
\def\ljk{\l_{3jk}}
\def\st{{\tilde t}}
\def\tchi{{\tilde \chi}}
\def\nino{{\tilde{\chi}^0_1}}
\def\chinop{{\tilde{\chi}^+_1}}

\def\chinopm{{\tilde{\chi}^\pm_1}}
\def \mtn { m_{\nino}}
\def \mtch { m_{\chinopm}}
\def\ifb{\text{fb$^{-1}$}}

\begin{document}

\title{New signatures and limits on $R$-parity violation from resonant squark production}
\author{Angelo Monteux}

\affiliation{New High Energy Theory Center, Department of Physics and Astronomy,
\\Rutgers University, Piscataway, NJ 08854, USA
}
\emailAdd{amonteux@physics.rutgers.edu  }

\abstract{
We discuss resonant squark production at the LHC via baryonic $R$-parity violating interactions. The cross section easily exceeds pair-production and a new set of signatures can be used to probe squarks, particularly stops. These include dijet resonances, same-sign top quarks and four-jet resonances with large $b$-jet multiplicities, as well as the possibility of displaced neutralino decays. We use publicly available searches at $\sqrt{s}=8$ TeV and first results from collisions at $\sqrt{s}=13$ TeV to set upper limits on $R$-parity violating couplings, with particular focus on simplified models with light stops and neutralinos. The exclusion reach of these signatures is comparable to $R$-parity-conserving searches, $m_{\tilde t}\simeq 500-700$ GeV. In addition, we find that O(1) couplings involving the stop can be excluded well into the multi-TeV range, and stress that searches for single- and pair-produced four-jet resonances will be necessary to exclude sub-TeV stops for a {\it natural} SUSY spectrum with light higgsinos.
}

\maketitle

\section{Introduction}
The Large Hadron Collider (LHC) has recently re-started collisions at a center of mass energy $\sqrt s=13\tev$, which will help to explore scenarios of New Physics well into the TeV range. This comes after a host of null results for supersymmetry (SUSY) with 8\tev\ collisions, which have given strong limits on colored superpartners (between 700\gev\ for the stop and 1500\gev\ for the gluino in \rpcing models). Although the majority of SUSY searches are based on the assumption of \rpa \ \cite{Farrar:1978xj} and on the missing energy signature of the lightest super-partner (LSP), the neutralino, the possibility of \rpv\ \cite{Hall:1983id,Bento:1987mu} has also been explored in great detail. The  common justification for imposing \rpa\ is to ensure proton stability, but the same result can be achieved by separately imposing the conservation of either baryon number or lepton number  (see also Ref. \cite{Faroughy:2014tfa} where a careful choice of symmetries ensures proton stability with both $B$ and $L$ violation). From a UV prospective, there is no strong argument for or against \rpa
\cite{Bento:1987mu,Ibanez:1991pr,Dreiner:2012ae}, and models can easily be built such that the low-energy effective theory has an accidental discrete symmetry resulting in proton stability.

In this work, we will be studying collider signatures of baryonic \rpv\ (RPV), where in addition to the field content and interactions of the Minimal Supersymmetric extension of the Standard Model (MSSM) the following super-potential operator is introduced:\footnote{Leptonic \rpv\ is relatively more constrained as the collider signatures include multiple hard leptons, see e.g. Refs.~\cite{Aad:2014iza,Chatrchyan:2013xsw,Graham:2014vya}. Limits from resonant slepton production have been discussed in \cite{Dreiner:2012np} in a spirit similar to the present work.}
\beq\label{eq:RPV}
W_{BRPV}=\frac{\l_{ijk}}2  U_i^c D_j^c D_k^c\,.
\eeq
The superfields $U_i^c, D^c_i,\ i=1,2,3$ contain the right-handed quarks and squarks (in the following, we use the two-spinor notation reviewed in Ref.~\cite{Dreiner:2008tw}) and the color indices have been contracted with the antisymmetric Levi-Civita tensor $\epsilon_{abc}$; gauge invariance enforces antisymmetry of the $\lijk$ coupling with respect to the exchange $j\leftrightarrow k$, $\lijk=-\l_{ikj}$.

In this work, we will study the experimental signatures of squarks resonantly produced via this operator. Before going ahead, we quickly review the implications of these additional interactions: first, they contribute to flavor-changing neutral currents (FCNCs), potentially contributing to flavor physics observables. There is a multitude of processes giving constraints on both individual RPV couplings and products of two couplings with different flavor indices \cite{Barbier:2004ez}: here it sufficient to point out that the stronger limits apply to couplings involving multiple first-generation couplings, while weaker limits involving second and third generation indices allow for larger couplings. In addition, all the limits become weaker as the relevant superpartner masses are increased. Such limits can be easily be accommodated by assuming particular flavor {\it ansatze}, in which the flavor structure of the Yukawas imposes hierarchical structures on the RPV couplings \cite{Nikolidakis:2007fc,Csaki:2011ge,Florez:2013mxa,Monteux:2013mna}.

For what concerns the LHC, \rpv\ removes the missing energy signature of \rpcing SUSY and replaces it by all-hadronic final states (possibly with multiple top quarks, depending on the flavor structure of RPV) \cite{Dreiner:1991pe,Allanach:2012vj,Durieux:2013uqa,Bhattacherjee:2013gr}. Both the ATLAS and CMS collaborations have performed multiple analyses in this scenario, with particular focus on pair-production of gluinos \cite{Aad:2014pda,Aad:2015rba,Chatrchyan:2013gia,CMS:2013qua}, followed by decays involving three or five jets, or to multiple same-sign top quarks resulting in same-sign dileptons with high $b$-multiplicities. Limits on gluinos are near or above  1\tev, with $\mtg\gtrsim1\tev$  confirmed by independent groups recasting the original searches to allow for more generic SUSY spectra \cite{Evans:2013jna,Graham:2014vya} (see also Ref.~\cite{Franceschini:2015pka} for a summary of the experimental results as of mid-2015). For what concerns squarks, the focus has been on pair-produced squarks decaying to two or four jets (possibly 3 jets+1 top), with limits on first- and second-generation squarks in the range $500-700\gev$ \cite{Durieux:2013uqa,Evans:2013jna} and the first limits on stops decaying to two jets in the range $350-400\gev$ presented in Refs. \cite{Khachatryan:2014lpa,ATLASCONF2015026}. Earlier limits on pair-produced stops resulting in same-sign leptons were shown in Ref.~\cite{Evans:2012bf}, based on collisions at $\sqrt s=7\tev$, although for a specific stop-neutralino splitting.

Finally, it is worth mentioning that \rpv\ has important consequences for baryogenesis: on one hand, if the baryon asymmetry is generated above the weak scale (more precisely, after the freeze-out of baryon-number-violating interactions mediated by squarks, at temperatures $T_{f.o.}\approx\mtq/20$) and {\it any} of the \rpving couplings is large, $\lijk\gtrsim 10^{-8}$, the baryon asymmetry is erased by these baryon-number-violating interactions \cite{Campbell:1991at,Dreiner:1991pe}. This is often rephrased by saying that baryogenesis implies small RPV couplings, so small that RPV decays are usually displaced at colliders \cite{Dreiner:1992vm,Barry:2013nva}. Because displaced decays are subject to much smaller backgrounds than prompt decays, the limits on displaced RPV are actually significantly more stringent than for prompt RPV, and seem to exclude any sub-TeV squarks \cite{Liu:2015bma,Csaki:2015uza,Zwane:2015bra}. What this argument overlooks is that, on the other hand, baryogenesis can be directly produced via \rpving interactions when some out-of-equilibrium particle $X$ decays well below the superpartner scale \cite{Dimopoulos:1987rk,Cline:1990bw,Mollerach:1991mu,Rompineve:2013xwa,Monteux:2014hua}. There are many working examples of this mechanism and in general the baryon asymmetry takes the form
\beq
Y_B= Y_X\times \frac{|\l|^2}{(8\pi)^\ell} \Phi_{CP}  \ f(M_X,\mt)\,,
\eeq
where $Y_i=n_i/s$ is the ratio of number density for the particle species $i$ and entropy density, $\ell$ is the number of loops at which the asymmetry is generated, $\Phi_{CP}$ is a CP-odd phase originating from the model parameters (e.g. soft masses or $A$-terms) and $f$ is a function of the various masses that depends on the kinematics of the process. To reproduce the observed value of the matter-antimatter asymmetry, $Y_B\simeq 10^{-10}$, one cannot have arbitrarily small RPV couplings. As an order-of-magnitude estimate, substituting $Y_X\lesssim 10^{-3}$ for a particle $X$ that was a thermal relic and taking the other coefficients to be at most O(1), one finds $\l\gtrsim 10^{-3}$. We will take this as hint that if \rpv\ has anything to do with baryogenesis, large couplings are preferred, while in the opposite case it should give displaced vertices and is already excluded below 1\tev; if large RPV couplings were to be excluded, baryogenesis via RPV would have to be generated from a non-thermal relic.


Large couplings give us another (so far unexplored) window to explore \rpv, as \textit{resonant production} of a single squark can be large enough to give observable signals at the LHC, with a cross section  potentially orders of magnitude larger than the one for pair-production of squarks and gluinos.\footnote{
Earlier works discussing colliders implications of resonant squark production include Refs.~\cite{Berger:1999zt,Berger:2000zk} proposing searches for the decay $\st\to b\chinop\to b W^+\nino$ at the Tevatron, and Ref.~\cite{Desai:2010sq} focusing on the $\st\to t\nino$ decay at the LHC. The results in the present work are the first direct limits based on experimental data.} In this work we will explore this possibility and find new signatures and limits, while also re-phrasing all the available limits in order to include the dependence on the RPV couplings and the SUSY spectrum. We will see that resonant production can probe squarks well above the reach of the already mentioned pair-production modes.

 This article is organized as follows: in Section \ref{sec:monosquark}, we recall the cross section for the resonant production of squarks through the \rpving\ interactions and we analyze the possible decay modes which depend on the superpartner mass spectrum. In section \ref{sec:stopLSP}, we show limits on a stop LSP coming from dijet resonances searches, while we present the phenomenologically richer scenario of a neutralino LSP in the following section \ref{sec:inoLSP}: here we discuss the same-sign top signature for both resonantly and pair-produced stops, the possibility of displaced neutralino decays, and the chargino-mediated decays. In Section \ref{sec:lhc13}, we discuss the prospects for discovery at Run 2 of the LHC and analyze the implications of a more generic supersymmetric spectrum: to cover blind spots of present analyses, we propose new searches that will better constrain \rpv. Conclusions are presented in Section \ref{sec:conclusions}.

\section{Single squark production: cross section and signatures}\label{sec:monosquark}
\subsection{Resonant production}\label{sec:monoprod}
In this section, we study the single production of a squark at the LHC. Because of the structure of the RPV operator in Eq. \eqref{eq:RPV}, only right-handed squarks can be resonantly produced through the scattering of quarks. For the third generation squarks, we will assume a purely right-handed stop $\st_1$, $\cos\theta_\st=1$; the cross section and the limits presented will get weaker for large left-right mixing, while in the case of a lightest stop which is mostly left-handed constraints will apply on the heavier stop $\st_2$.

At the parton level and at leading-order the cross section is \cite{Dimopoulos:1988jw}:\footnote{Note that the different normalization of the RPV operator in Eq. \eqref{eq:RPV} (choice consistent with the one in the review \cite{Barbier:2004ez}) translates in a different overall factor than in Refs. \cite{Dimopoulos:1988jw,Berger:1999zt,Berger:2000zk,Plehn:2000be}.}
\beq\label{eq:monocrossx}
\hat \sigma (d_j d_k\to \tilde u_i^*)=\frac{\pi}6|\lijk|^2\frac{\cos^2\theta_{\tilde u_i}}{m_{\tilde u_i}^2}\delta(1-m_{\tilde u_i}^2/ \hat s)\,.
\eeq
At the LHC, the cross-section is obtained by convoluting $\hat\sigma$ with the proton parton distribution functions (PDF) for the relevant quarks in the initial state. We used the \texttt{NNPDF2.3} set with QED corrections \cite{Ball:2013hta} and the \texttt{ManeParse} toolbox \cite{Godat:2015xqa} to manipulate the PDFs; for this set, the uncertainties are of the order of 5\%. The numerical results match perfectly the ones computed with the MC generator \texttt{MadGraph5} \cite{Alwall:2014hca}, with UFO model files for the RPV interactions generated with \texttt{Feynrules 2.0} \cite{Alloul:2013bka,Fuks:2012im}.
The next-to-leading order (NLO) cross section including QCD contributions was computed in Ref. \cite{Plehn:2000be}, where it was found that the LO cross section is increased by a factor of order $1.2-1.3$, depending on the flavor indices of the RPV coupling as well as the mass of the squark. The uncertainties due to the renormalization and factorization scale dependence were also computed and found to be less than $5\%$. We include these $K$-factors in the cross section; by combining the PDF uncertainties with the renormalization and factorization scale dependence, the theory error on the cross section is estimated to be smaller than about $7\%$.

\begin{figure}[t]
\begin{center}
\includegraphics[width=0.48\textwidth]{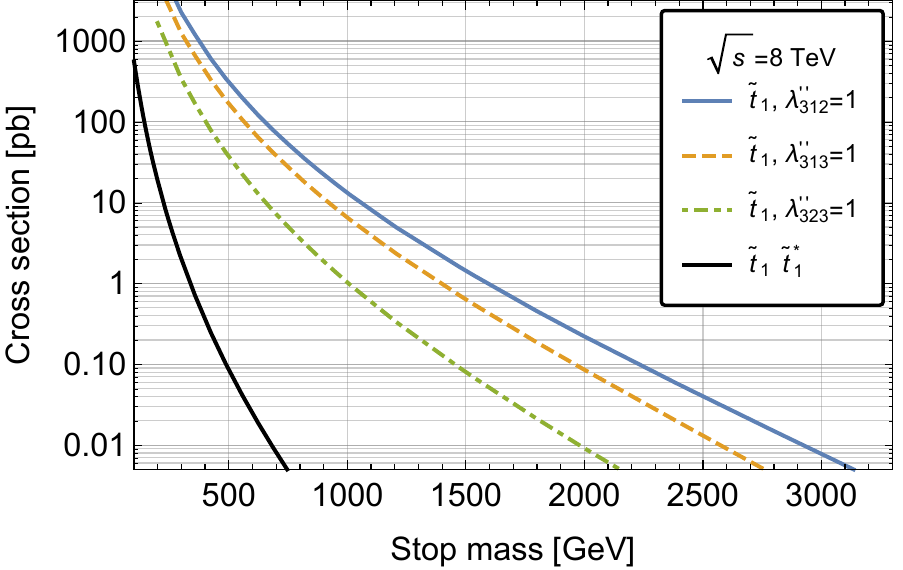}\hfill
\includegraphics[width=0.48\textwidth]{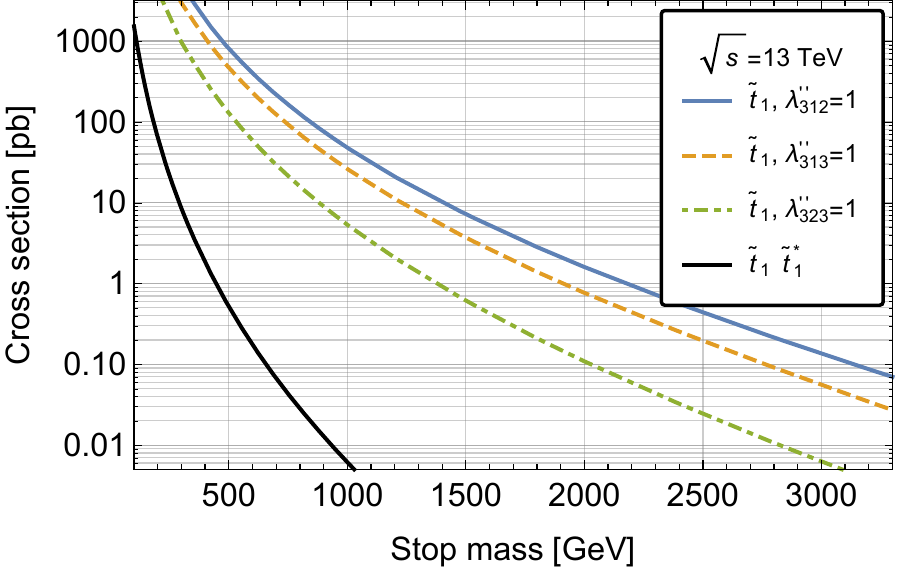}
\caption{Single stop production cross section for proton-proton collisions at $\sqrt s=8\tev$ (left) and $13\tev$ (right) as a function of the right-handed stop mass, with a single RPV coupling fixed to one and including NLO effects. For smaller couplings, the cross section scales as $\lambda_{3jk}^{''2}$. We present the sum of $\st$ and $\st^*$ cross sections. The NLO cross section for stop pair-production is shown for reference. 
}
\label{fig:monocrossx8}
\end{center}
\end{figure}

For collisions at the LHC with a center of mass energy of 8 and 13 \tev, the cross sections for the production of a right-handed stop is shown in Fig. \ref{fig:monocrossx8}, where for reference the relevant RPV couplings have been set to unity in each case (in addition, we have summed the cross sections for stop and anti-stop production, see more details below). For comparison, the cross section for (\rpa-conserving) stop pair-production is also depicted: even for smaller couplings $\ljk<1$, resonant production can be orders of magnitude more efficient than pair-production, particularly in the multi-TeV region. Here and in the following, we turn on one production channel at a time by assuming that the corresponding RPV coupling is dominant, and then find limits on that specific coupling within this assumption. Having several large \rpving couplings would increase the cross section but the (potentially stronger) corresponding limits would be on linear combinations of couplings (squared). In this sense, the limits presented in this work will be conservative.

\begin{table}[t]
\begin{center}
\begin{tabular}{|c|ccccccccc|}\hline
$\mtt$ [GeV] & 200   & 500 &  1000 &  1500 & 2000 & 2500  & 3000 &  3500 &4000\\\hline
$\sigma (\bar d\bar s\to \st\,)/\sigma (d s\to \st^*)$ 
& 0.38 & 0.27 & 0.18 & 0.13 & 0.098 & 0.075 & 0.058 & 0.046 & 0.039 \\
$\sigma (\bar d\bar b\to \st\,)/\sigma (d b\to \st^*)$ 
& 0.36 & 0.25 & 0.16 & 0.12 & 0.088 & 0.066 & 0.052 & 0.045 & 0.046 \\
$\sigma (\bar b\bar s\to \st\,)/\sigma (b s\to \st^*)$ 
& 1.0 & 0.99 & 0.99 & 0.98 & 0.96 & 0.94 & 0.90 & 0.85 & 0.80 \\\hline
\end{tabular}
\caption{Ratio of stop and anti-stop production rates, $\sigma (\bar d_j\bar d_k\to \st\,)/\sigma (d_j d_k\to \st^*)$, at the LHC with $\sqrt s=8\tev$, due to different parton luminosities. For $\sqrt s=13\tev$, the ratios are typically larger, especially at large masses.}
\label{tab:ststbar}
\end{center}
\end{table}%

In this work, we will focus on the resonant production of stops: the couplings $\ljk$ are less constrained by flavor-changing neutral currents and flavor physics \cite{Barbier:2004ez}, while at the same time they are expected to be the largest in models where flavor symmetries determine a hierarchical structure of the RPV couplings \cite{Monteux:2013mna,Nikolidakis:2007fc,Csaki:2011ge,Florez:2013mxa}.  We note that the cross section for any (right-handed) up-type squark $\tilde u_i$ is the same as for the stop, with the substitution $\ljk\to\lijk$. Down type squarks could also be produced resonantly, but given the absence of top quarks in the proton PDF, the only couplings that can be probed are $\lijk, i\neq3$. Couplings involving multiple first-generation quarks are constrained by flavor physics to be well below one \cite{Barbier:2004ez}, so we do not investigate production via $\l_{1jk}$.  The couplings $\l_{2jk}$ are relatively less constrained and we will shortly discuss their signatures in Section \ref{sec:lhc13}, but the cross section will be somewhat suppressed by the charm quark PDF.

A particular feature of this channel is that production of anti-stops via the couplings $\l_{312}$ and $\l_{313}$  makes use of valence $d$ quarks, while production of stops needs virtual anti-quarks; thus, there is an asymmetry between the yield of squarks and anti-squarks. For $\sqrt s=8\tev$, the ratio of cross sections $\sigma (\bar d_j\bar d_k\to \st\,)/\sigma (d_j d_k\to \st^*)$  is in the range $0.1-0.4$ (as listed in Table \ref{tab:ststbar}) for stop masses in the range $200\gev-4\tev$. As expected, anti-stop production prevails for $\l_{312},\l_{313}$ while the ratio for resonant production via $\l_{323}$ is nearly one.

\begin{figure}[t]
\begin{center}
\subfloat[]{\includegraphics[height=0.11\textheight]{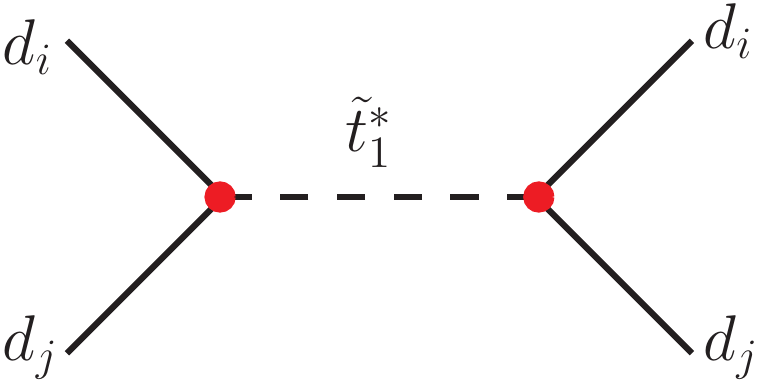}}\hfill
\subfloat[]{\includegraphics[height=0.11\textheight]{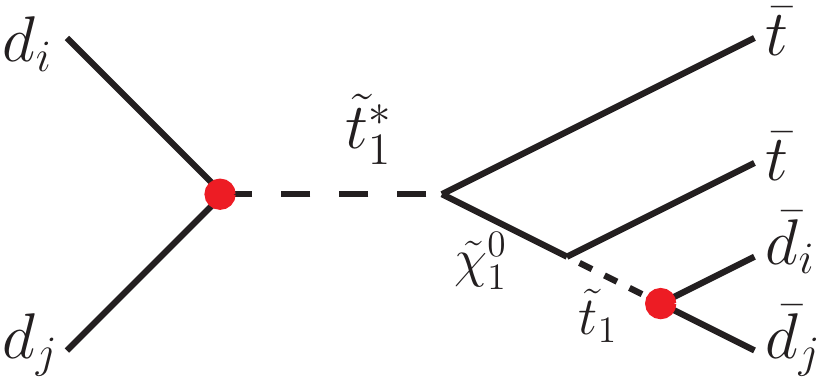}}\hfill
\subfloat[]{\includegraphics[height=0.11\textheight]{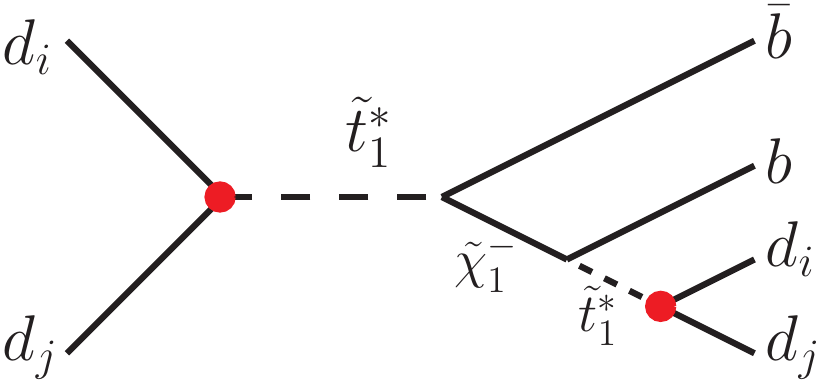}}
\caption{Stop resonant production and decay modes considered in this work: (a) dijet resonances, (b) same-sign tops  and (c) $b$-rich multi-jet resonances.  Red dots denote the \rpving couplings, $\ljk$. The respective conjugated processes are not shown.}
\label{fig:decaymodes}
\end{center}
\end{figure}

\subsection{Stop decays and signatures}\label{sec:monodecay}
The decay modes of the stop will in general depend on the detailed superpartner mass spectrum and they will include both \rpcing and \rpving channels. The RPV decay $\st\to d_j d_k$ is always present and corresponds to a dijet resonance, while the \rpcing modes depend on which supersymmetric states are lighter than the stop: for simplicity we will study simplified models where the only light states are the stop and a neutralino (including charginos if appropriate). For concreteness, we decouple all the other superpartners by setting their mass to 10\tev. 

If the stop is the Lightest Supersymmetric Particle (LSP), its decay via off-shell neutralinos or gluinos will always be subdominant compared to the direct decay $\st\to jj$. For a non-LSP stop, the decay modes to neutralinos/charginos will open; the dijet decays will dominate over other decay channels for large RPV couplings, $\ljk\gg0.1$, with the only possible exception being the decay to a gluino, $\st\to t \tg$ (forbidden in the simplified spectrum). As limits on RPV gluinos are in the range $1-1.5\tev$ depending on the SUSY spectrum \cite{Evans:2013jna,Graham:2014vya,Chatrchyan:2013gia,Chatrchyan:2013fea,Aad:2015lea,Aad:2014pda}, the simplified models with a neutralino LSP will be good proxies for a more complex spectrum in the range $\mtt\lesssim1\tev$. 

In supersymmetric extensions of the Standard Model, neutralinos are admixtures of the available neutral fermions, which here are the bino, the wino and two higgsinos. In general, the neutralino mixing matrix $N_{ij}$ depends on the supersymmetric $\mu$ term and on the soft SUSY breaking parameters $M_1,M_2$, with three asymptotic scenarios dictating the experimental signatures:
\begin{itemize}
\item bino-like $\nino$: for $M_1\ll\mu,M_2$, the lightest neutralino $\nino$ is a pure bino and it is decoupled from all the other neutralinos and charginos,  $\mtn\ll m_{\tchi^0_{2,3,4}}, m_{\tchi^\pm_{1,2}}$. The decay mode $\st\to t \nino$ is accessible for $\mtt>m_t+\mtn$, while at lower masses the decay rate with an off-shell top ($\st\to b W^+\nino$) decreases quickly. As the neutralino is also unstable, $\nino\to tjj,\bar tjj$, the final state of  a stop decay contains two tops, which can be same-sign.

\item higgsino-like $\nino$: for $\mu\ll M_1,M_2$, the light spectrum includes two almost-degenerate neutral states $\tchi^0_{1,2}$ and a chargino $\tchi^+_1$, with the other states decoupled, $m_{\tchi^0_{1,2}} \simeq\mtch \ll m_{\tchi^0_{3,4}}, m_{\tchi^\pm_{2}}$. Here the decay $\st\to t\tchi^0_{1,2}$ is always subdominant (although the branching ratio can be sizable, up to $O(1/2)$) with respect to the chargino decay $\st\to b\chinop$, which is not phase-space suppressed. The chargino is also unstable and for each stop there is  a $b$-rich final state $\st\to bb jj$, where, depending on the relevant RPV coupling, another one of the final jets might be a $b$-jet. This corresponds to a four-jet resonance, with a three-jets sub-resonance inside.

\item wino-like $\nino$: for $M_2\ll \mu, M_1$, the lightest states are a neutralino and two charginos, $\mtn\simeq \mtch \ll m_{\tchi^0_{2,3,4}}, m_{\tchi^\pm_{2}}$. As the focus of this work is on purely right-handed squarks, their interactions with a wino are vanishing and we will not consider this scenario in details. For a squark with appreciable left-right mixing, the resonant cross section will be smaller, but the conclusions will be qualitatively similar to the other two cases (particularly to the higgsino-like case), as both previous decay modes $\st\to t\nino$ and $\st\to b\chinop$ will be available.
\end{itemize}

To summarize, the signatures of resonant stop production are dijet resonances, same-sign tops (accompanied by two extra jets) and four-jet resonances containing a three-jet sub-resonances and many $b$ jets. The last two signatures depend on the nature of the LSP and will both be present in the case of a well-mixed neutralino. As anti-stops are preferably produced, same-sign anti-tops would dominate the signal (unless the relevant RPV coupling is $\l_{323}$ for which $tt$ and $\bar t\bar t$ are produced at nearly the same rate).

Finally, we emphasize that in the case of resonant production with decay to a non-dijet final state, the limits will not decouple for large $\l$, even though the branching ratios for decays other than the production channel become small: this is because the signal yield is $\sigma\times Br_i$, and one can schematically write
\beq\label{eq:nondecoupling}
\sigma\times Br_i\ \propto\  \Gamma_{q q} \times \frac{\Gamma_i}{\Gamma_{q q}+\Gamma_i}\propto\frac{\l^2g^2}{\l^2+g^2}
\sim 
\left\{\begin{array}{ll} g^2 &\text{for } \l\gg g \\ \l^2 &\text{for } \l\ll g\end{array}\right.\ ,
\eeq
where $g$ is a coupling relevant for the non-dijet decay channel and can be redefined to include numerical coefficients and phase-space suppression. 

The same-sign top signature of RPV SUSY was first discussed in Ref. \cite{Durieux:2013uqa}, where limits on gluinos and first- and second-generation squarks were found within the assumption of a particular (minimal flavor violating) structure of the RPV couplings. Resonant production was subdominant and same-sign tops came from pair-produced squarks, each giving one top in the final state after the neutralino decay. This work discusses a new signature as the decay of a single stop can by itself give same-sign tops (before the LHC, this was discussed in Ref.~\cite{Desai:2010sq}). In addition to studying resonant stop production according to the signatures outlined above, in the next sections we will also examine novel signatures from stop pair-production, such as four-top final states.

We conclude this section by noting that, within this set of assumptions, the usual flavor physics limits on \rpv\ \cite{Barbier:2004ez} are rather weak, as those preferably involve either first- and second-generation fields, multiple RPV couplings and/or non-negligible mixing in the squark sector. In the hypothesis of a single dominant coupling $\l_{3jk}$, the most relevant process  is $n-\bar n$ oscillation caused by a one-loop box diagram with exchange of $W$ boson - chargino as well as quark - squark pairs, as first proposed in Ref.~\cite{Chang:1996sw} and recently reviewed in \cite{Calibbi:2016ukt}. The largest diagram involves insertions of top and bottom quark masses, together with left-right (LR) squark mixing (for both the stop and the sbottom). Within the simplified models discussed above, this would be relevant for a wino LSP  with maximal LR squark mixing. The resulting limits are $\l_{312},\l_{313}\lesssim 0.1 (\mtt/400\gev)^2$, although the uncertainties for the six quark hadronic matrix element are large and can relax the limit by a factor of a few. Improvements on the $n-\bar n$ oscillation lifetime could be achieved by a proposed experiment at the European Spallation Source (ESS) in Lund, Sweden. We refer to the recent study \cite{Calibbi:2016ukt} for a detailed discussion of $n-\bar n$ oscillation and prospects for future improvements.

\section{Results: stop LSP}\label{sec:stopLSP}
In this section we study limits on resonant stop production set by dijet resonances searches with the full LHC dataset at $\sqrt s=8\tev$ and with first results at 13\tev. We have checked all publicly available dijet resonance searches, including searches at the Tevatron, and we find that the strongest limits are set by:
\begin{itemize}
\item the ATLAS search for a Gaussian dijet resonance based on 20.3 \ifb\ of collisions at $\sqrt s=8\tev$ \cite{Aad:2014aqa}, which studies a mass region between 300\gev\ and 4.2\tev, as well as preliminary results based on $3.6\ \ifb$ at $\sqrt s=13\tev$ \cite{ATLAS:2015nsi}. The limits are given as 95\% C.L. upper limits on $\sigma\times A \times Br$, for given values of the width of the Gaussian resonance. The acceptance $A$ is given by the fraction of events that passed the kinematic cuts and are near the Gaussian peak in the dijet invariant mass distribution, as explained in Appendix \ref{sec:accept}, and varies between $50\%$ for the 8\tev\ analysis at $\mtt=300\gev$ and $17\%$ for the 13\tev\ analysis at $\mtt=4\tev$.

\item the CMS wide dijet searches based on 19.8 \ifb \ and 18.8 \ifb\ of collisions at $\sqrt s=8\tev$ \cite{Khachatryan:2015sja,CMS-PAS-EXO-14-005} as well as preliminary results based on $2.4\ \ifb$ at $\sqrt s=13\tev$ \cite{Khachatryan:2015dcf}. In these searches the whole dijet mass distribution is used so that the acceptance for such searches is higher and varies between $50\%$ and $16\%$ for the 8\tev\ dataset, see Appendix \ref{sec:accept}. Ref. \cite{CMS-PAS-EXO-14-005} is based on novel {\it scouting} techniques, where after requiring loose kinematic cuts, only data about the reconstructed jets (such as four-vectors) is stored while the full detector-level event is erased. This allows to cover low-mass regions where the QCD background would otherwise be overwhelming due to high event rates.
\end{itemize}

Before we discuss limits from  resonant stop production, we also recall the LHC searches for pair-produced stops resulting in two  dijet resonances. In Ref. \cite{Khachatryan:2014lpa}, the CMS collaboration investigated this scenario, distinguishing between final states involving one or zero $b$-quarks. For $\st\to q q$ ($\st\to b q$), stop masses were excluded between 200\gev{} and 350\gev\; (resp. 385\gev). In a similar search, ATLAS   was able to exclude pair-produced stops with $\st\to b s$ for stop masses between 100\gev\ and 310\gev\, \cite{ATLASCONF2015026}, thus also excluding the lower mass range that was not covered by the CMS search. In both cases, a right-handed stop LSP was assumed for the branching ratio to jets to be one.

\begin{figure}[t]
\begin{center}
\includegraphics[width=0.5\textwidth]{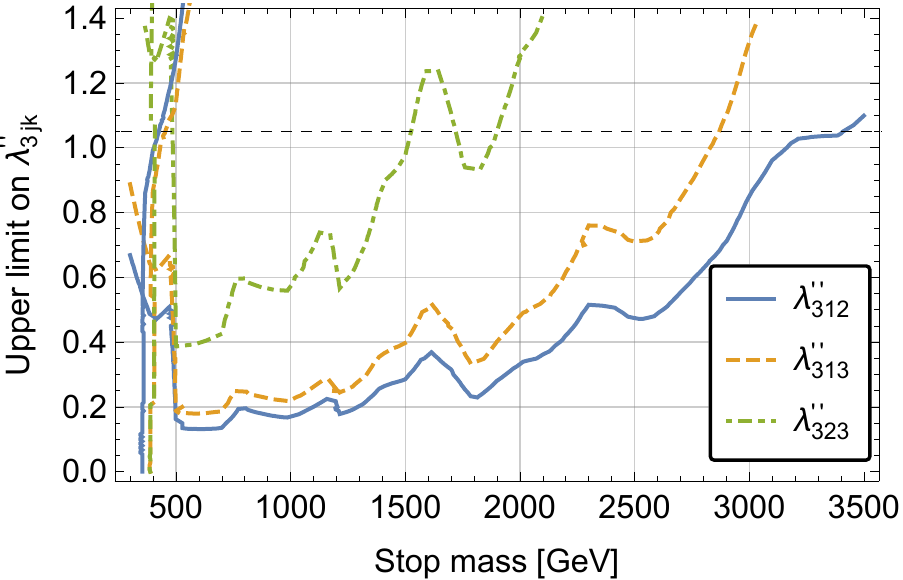}
\caption{Upper limits on $\ljk$ from the ATLAS and CMS dijet resonant searches discussed in the text, as a function of the stop LSP mass. Couplings above each line are excluded. The dashed horizontal line shows the perturbativity limit, $\ljk<1.05$ \cite{Allanach:1999mh}. The region to the left of the nearly vertical lines at 350\gev\  and 400\gev\ is excluded by searches for pair-produced dijet resonances.}
\label{fig:stopLSP}
\end{center}
\end{figure}

We show the combined results from all dijet searches in Fig. \ref{fig:stopLSP}, where we find upper limits on each \rpving coupling $\ljk$ as a function of the stop mass, assuming that only one coupling is responsible for single production of squarks.\footnote{
The signal will increase if multiple RPV couplings are allowed to be  large, but dangerous contributions to FCNCs would arise \cite{Barbier:2004ez}, in particular to kaon mixing \cite{Slavich:2000xm} and exotic $B$ decays. Allowing for multiple RPV couplings, we find that the direct limits in Fig.~\ref{fig:stopLSP} give stronger constraints than flavor physics for all the individual  couplings $\l_{312},\l_{313},\l_{323}$, as well as for the product of couplings $|\l_{313}\l_{312}|$, while the products $|\l_{312}\l_{323}|$ and $|\l_{313}\l_{323}|$ respectively receive stronger constraints from $B^-\to\phi^0\pi^-$ and $K-\bar K$ (by factors of about 4 and 10). In any case, even a mild hierarchy between different couplings can nullify all flavor constraints while still allowing one O(1) coupling. This can now be directly excluded by the results for resonant stop production in this work.
}
 The solid blue line gives the limits on $\l_{312}$, while the dashed orange and the dot-dashed green lines are the limits on $\l_{313}$ and $\l_{323}$, respectively. The {\it scouting} analysis \cite{CMS-PAS-EXO-14-005} gives the strongest limits between 500\gev\ and 1\tev. The first limits from the 13\tev \ dataset are included and they already dominate at large masses, $\mtt\gtrsim 1.2\tev$.  The regions to the left of the nearly vertical lines between 350\gev\  and 400\gev\ are excluded by searches for pair-produced dijet resonances. A dashed horizontal line shows the upper limit  $\ljk<1.05$ that is found by considering RGE evolution of the RPV coupling and by requiring the couplings to be perturbative up to the GUT scale \cite{Allanach:1999mh}. The direct LHC limits are stronger than the perturbativity bound for stop masses  up to between 1.5 and 3.3\tev, depending on the flavor structure of the RPV coupling. At low masses, $\ljk=1$ is excluded for all couplings except for $\l_{323}$, where  the narrow range $400\gev <\mtt<500\gev$ is still allowed. This narrow window is also the only region where our limits can be weaker than those from $n-\bar n$ oscillation discussed earlier (those are not shown as they rely on additional assumptions such as LR mixing and the wino mass). It should be noted that this low-mass range is not excluded by previous Tevatron data: upper limits on dijet production cross sections from the CDF collaboration \cite{Aaltonen:2008dn} are larger than the RPV resonant cross section for $p\bar p$ collisions at $\sqrt s=1.96\tev$ even for $\ljk=1$.

When using these results on pair-produced stops, we have also included another production channel that so far had not been discussed in the literature: the scattering process $d_j\bar d_j\to \st \st^* $ via two insertions of the RPV operator $\ljk$ and the $t$-channel exchange of a  $d_k$ quark. For O(1) couplings, the cross section can be sizable. Without considering interference with other QCD pair-production modes, the cross section will be proportional to $\l^4_{3jk}$, while interference gives terms proportional to $g_S^2\l^2_{3jk}$. This can increase the cross section by a factor of order 1 for $\ljk=1$. We include this effect in computing our limits and see that it is responsible for extending the paired-dijets resonance limits to slightly higher values of the stop mass for O(1) RPV couplings (e.g. 450\gev\ instead of 380\gev). The net effect is negligible as these regions are independently excluded by the resonant dijet searches.

Although in this section we have assumed that dijets are the only decay channel of the stop, it is expected that the limits found will be fairly robust with respect to adding other light superpartners to the spectrum: because the excluded regions correspond to large values of the couplings ($\ljk\gtrsim 0.2$) and this channel is never phase-space suppressed, for the stop to have sizable branching ratios into other superpartners requires large couplings, as well as decays into light final states. Lighter neutralinos will not change these limits by much unless they are higgsino-like, in which case the decay to chargino can be sizable due to the top Yukawa coupling. We will investigate these scenarios in the next section. As gluinos below 1\tev\ are excluded by other \rpving searches, dijet decays will dominate at large $\ljk$ for stops up to about 1.5\tev\ even if the gluino mass is 1\tev.

\section{Results: neutralino LSP}\label{sec:inoLSP}
In this section we study the signatures of the stop+neutralino simplified models described in Section \ref{sec:monodecay}. The stop branching ratios are defined as
\beq
Br_i=\Gamma_i/\Gamma_{tot},\qquad
\Gamma_{tot}=\Gamma(\st_1\to \bar d_j \bar d_k)+n_{\nino}\Gamma(\st_1\to t \nino)+n_{\chinopm}\Gamma(\st_1\to b \chinop)\,,
\eeq
where the coefficients $n_i$ count the number of light -inos, $ (n_\nino,n_\chinopm)=(1,0)$ for a bino-like LSP and $ (n_\nino,n_\chinopm)=(2,1)$ for a higgsino-like LSP. For reference, the analytical expressions for the decay rates of the stop to various final states are written in Appendix \ref{sec:decayrates}. In Fig. \ref{fig:br}, we set the neutralino mass to 200\gev\ and the RPV coupling to $0.1$ and display the dependence of the branching ratios on the stop mass. The phase-space suppression of the $\st\to t\nino$ decay can be easily identified, as well as the dominance of the chargino decay for the case of an higgsino-like LSP.

The first effect of adding more decay modes is to relax the limits from (single- and pair-produced) dijet resonances, due to the decreased branching ratios into dijet; nevertheless, we do not expect limits from single dijet resonance to decrease by much, while for pair-production, with limits on stop masses in the range 350-400\gev, we do not expect much change in the bino-like LSP model, as the branching ratio to $t\nino$ becomes sizable only for $\mtt\gtrsim \mtn+m_t\gtrsim 300\gev$. For the higgsino model, we expect the pair-produced dijet resonance limits to be present only at large $\l$ as the chargino decays easily prevail elsewhere.

\begin{figure}[t]
\begin{center}
\includegraphics[width=0.45\textwidth]{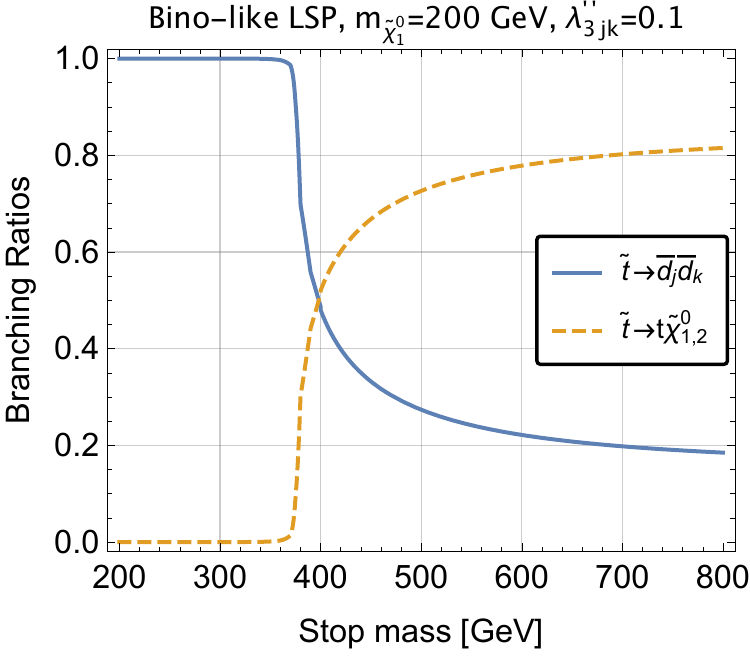}\qquad
\includegraphics[width=0.45\textwidth]{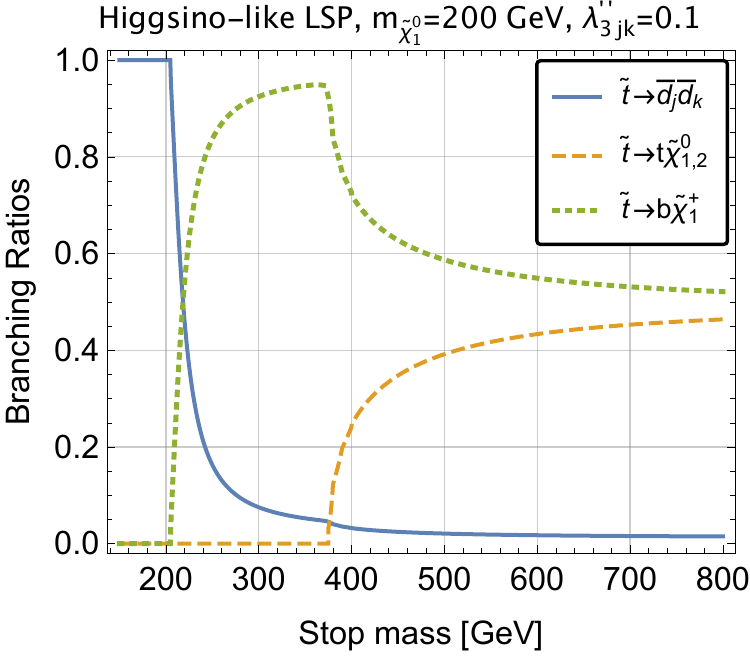}
\caption{Branching ratios of the stop as a function of its mass, for a bino-like (left) and a higgsino-like (right) LSP neutralino, with the \rpving coupling taken as $\ljk=0.1$ and the neutralino mass set to 200\gev. For larger $\ljk$, the dijet decay rate increases. For the higgsino case, the dashed orange line represents the sum of decay rates to both $\nino$ and $\tchi^0_2$, which are nearly degenerate.}
\label{fig:br}
\end{center}
\end{figure}

Finally, the neutralino and the chargino will also decay via the \rpving interactions and an off-shell stop (the analytical expressions for the decay rates are given in Appendix \ref{sec:decayrates}). For most of the parameter space, this decay will be prompt, apart from the case $\ljk\ll1$ and/or $\mtn\ll m_t$, where the final state includes an off-shell top. We will first discuss the prompt decays and defer the case with long-lived or collider-stable neutralinos to Section \ref{sec:dv}.

\subsection{Same-sign tops}\label{sec:SSt}
If the decay rate $\st\to t\nino$ is sizable, after the neutralino decay $\nino\to tjj,\bar tjj$ the final state will include same-sign (SS) tops, which in turn yield same-sign leptons. The most relevant searches at the LHC are the CMS same-sign lepton search \cite{Chatrchyan:2013fea} and the ATLAS four-top search \cite{Aad:2015kqa}, giving the 95\% C.L. limits on $tt$, $\bar t\bar t$ and $tt\bar t\bar t$ production in Table \ref{tab:cmsSS}. For the ATLAS search \cite{Aad:2015kqa}, we used the limits on sgluon production as the kinematics of the outgoing tops will be most similar to our model (especially for light neutralinos, the extra jets in $\nino\to t jj$ are soft).
One can see that there is some tension between expected and observed limits in the  CMS search \cite{Chatrchyan:2013fea}
. Although the model in  consideration could explain this excess (in particular, the production cross section of anti-tops can be higher than the cross section for tops and give a signal only in that channel), in this work we take a conservative approach and simply show excluded regions in the parameter space, keeping in mind that the observed limits are weaker than the expected limits by about $1.5$ standard deviations in the $\bar t \bar t$ channel.

\begin{table}[t]
\begin{tabular}{|c|c|c|c|c|}\hline
Process  & pp$\to tt$ \cite{Chatrchyan:2013fea} & pp$\to \bar t \bar t$ \cite{Chatrchyan:2013fea} & pp$\to tt\, \bar t \bar t$ \cite{Chatrchyan:2013fea} &  pp$\to tt\, \bar t \bar t$ \cite{Aad:2015kqa} \\\hline
Observed upper limit on $\sigma$ [fb] & 370 & 350 & 49 & 180 $-$ 18\\ 
Expected upper limit on $\sigma$ [fb] & $310^{+110}_{-80}$ & $160^{+140}_{-80}$ & $36^{+16}_{-9}$ & $150^{+50}_{-40} - 20^{+8}_{-5}$
\\\hline
\end{tabular}
\caption{Experimental upper limits on same-sign top production, from Refs. \cite{Chatrchyan:2013fea,Aad:2015kqa}. In the last column (from Ref.~\cite{Aad:2015kqa}) we listed the cross sections at 350\gev \ and 600\gev.}
\label{tab:cmsSS}
\end{table}%

The upper limits of Table \ref{tab:cmsSS} can be used to set limits on the following processes:
\beq\nn
&\text{resonant production}: \qquad pp\to \st \to t\nino\to t\,t d_j d_k +  t\,\bar t \bar d_j \bar d_k\,;\\\nn
&\text{pair-production:}\qquad pp\to \st\st^* \to (t\nino )(\bar t \nino)\to (t\,t d_j d_k)(\bar t\, \bar t \bar d_j \bar d_k) +  (t\,\bar t \bar d_j \bar d_k)(\bar t\, \bar t \bar d_j \bar d_k)\\\nn
&\hspace{8cm}+  (t\,t d_j d_k)(\bar t\, t  d_j d_k)+  (t\,\bar t \bar d_j \bar d_k)(\bar t\,  t  d_j d_k)\,,
\eeq
where in the first line the charge-conjugated process with anti-stop production is also present, and in each line both neutralino decay modes are equally probable,  $Br(\nino\to t d_j d_k)=Br(\nino\to \bar t\bar d_j\bar d_k)=\frac12$. While resonant production gives rise to same-sign di-top production (preferably $\bar t\bar t$), pair-production gives four tops, with up to three with a common sign. In this scenario, we can use both the SS di-top limits and the $tt\bar t \bar t$ limits in Table \ref{tab:cmsSS}.

\begin{figure}[t]
\begin{center}
\includegraphics[width=\textwidth]{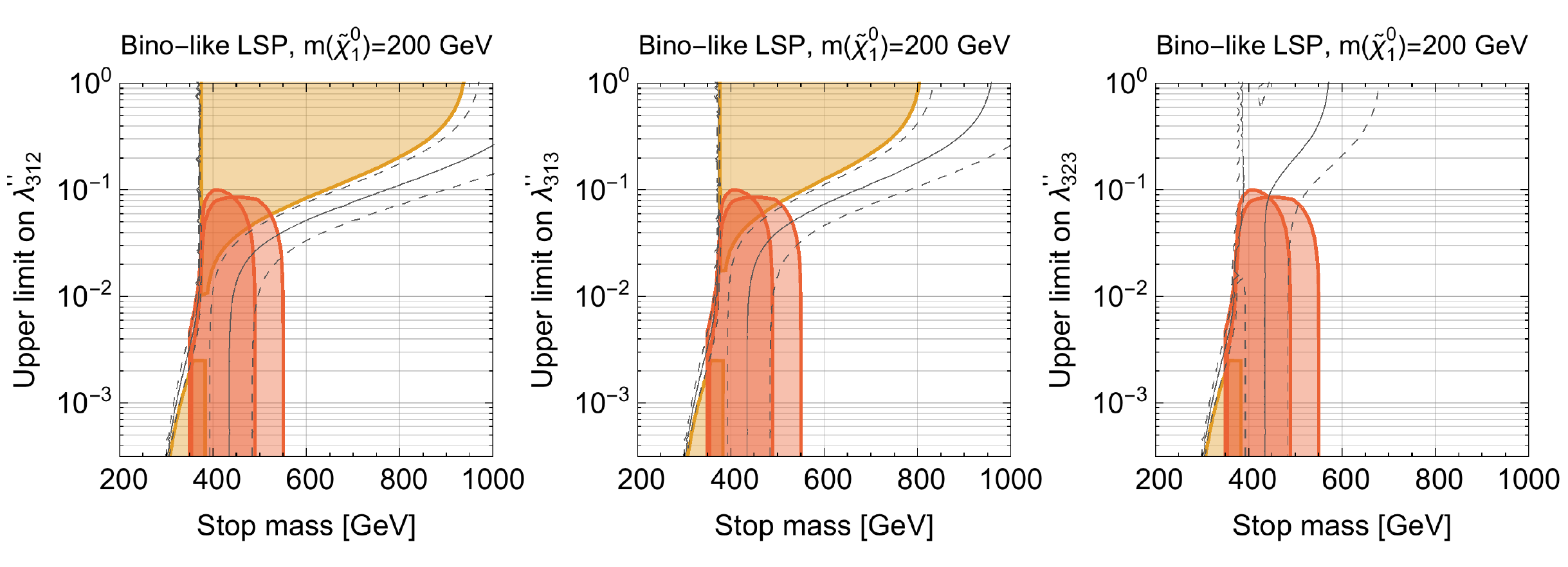}
\includegraphics[width=\textwidth]{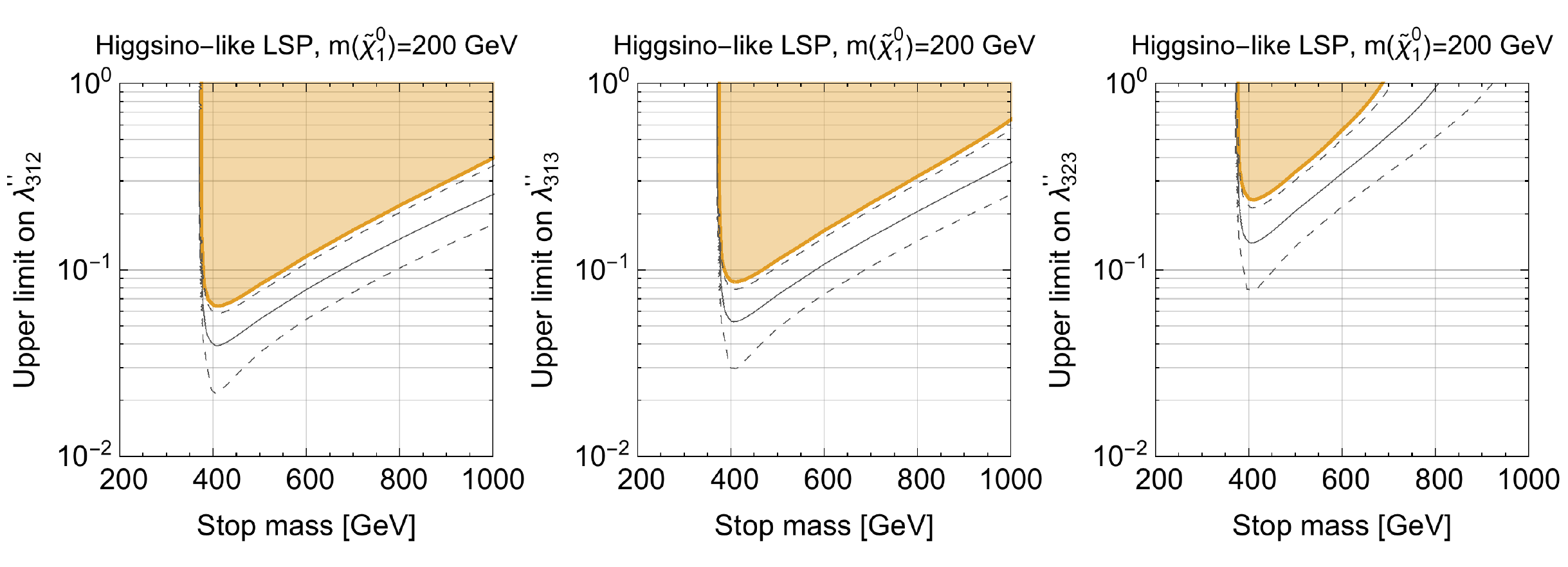}
\caption{Limits on RPV stops from same-sign top searches, specifically $pp\to \bar t\bar t$, and $ppp\to tt\,\bar t\bar t$ from the CMS search \cite{Chatrchyan:2013fea} based on a luminosity of 19.5 \ifb\ and the ATLAS search \cite{Aad:2015kqa} \ based on $20.3\ \ifb$, both at $\sqrt s=8\tev$. The neutralino mass is fixed  to 200\gev. The shaded regions show the observed exclusions, with regions excluded by $\bar t\bar t$ ($tt\,\bar t\bar t$) in orange (red: here we separately show the ATLAS and CMS exclusions, with the ATLAS results extending at larger masses); the expected limits on $\bar t \bar t$ are shown by gray solid lines (with their $1\sigma$ range delimited by dashed gray lines). Note the different range on the vertical axes between the top and bottom figures.}
\label{fig:SSinos}
\end{center}
\end{figure}

We show the resulting limits  on each RPV coupling $\ljk$ in Fig. \ref{fig:SSinos}, for either a bino-like LSP and a higgsino-like LSP; in the shaded regions the signal exceeds the observed limits, while the expected limits with relative uncertainties are shown as gray lines. Orange regions are excluded by $\bar t \bar t$ searches while red regions are exclude by $tt\,\bar t\bar t$. For large $\ljk$, the limits come from resonant stop production and subsequent decay into SS di-tops; because limits on production of $tt$ and on $\bar t\bar t$ are almost the same while resonant production preferably gives $\bar t\bar t$, the excluded regions are predominantly  driven by anti-stop production. For small RPV coupling, the limits come from stop pair-production and decays into SS di-tops as well as four tops ($tt\bar t\bar t$), with the limits from the second signal dominating over the first one. We note that resonant stop production via $\l_{323}$  is not constrained by the SS top signature in the bino-like LSP case, and only receives weak limits in the higgsino-like LSP case. This is due to both the smaller parton luminosities of $b$ and $s$ and to the observed limits being weaker than expected. Finally, the excluded regions behave as expected: for a bino-like LSP, the branching ratio for $\st\to t\nino$ dominates for $\ljk\lesssim 0.1$, and for even smaller couplings it can be sizable in the phase-space suppressed region $\mtt<\mtn+m_t$; on the other hand, with a higgsino-like LSP the unsuppressed branching ratio for $\st\to b\chinop$ dominates as soon as $\ljk\lesssim 0.1$, and SS top decays of pair-produced stops give no limits.

We conclude this section by noting that the limits quoted could be already improved by re-analyzing the LHC dataset with $\sqrt s=8\tev$: in fact, the signal from single stop production is not only SS tops, but also includes two accompanying jets for each stop (one of which could be a $b$ jet). While for $\mtn\simeq m_t$ these extra jets are soft and not easy to distinguish from the QCD background, for larger neutralino masses they could be hard enough to pass the cuts (see Sec.~\ref{sec:new} for more details). If a significant excess in SS tops was to be confirmed, searching for these extra jets would be necessary to confirm the RPV origin of the signal. A discussion with proposed cuts for this scenario can also be found in Ref. \cite{Desai:2010sq}.

\begin{figure}[t]
\begin{center}
\includegraphics[width=0.48\textwidth]{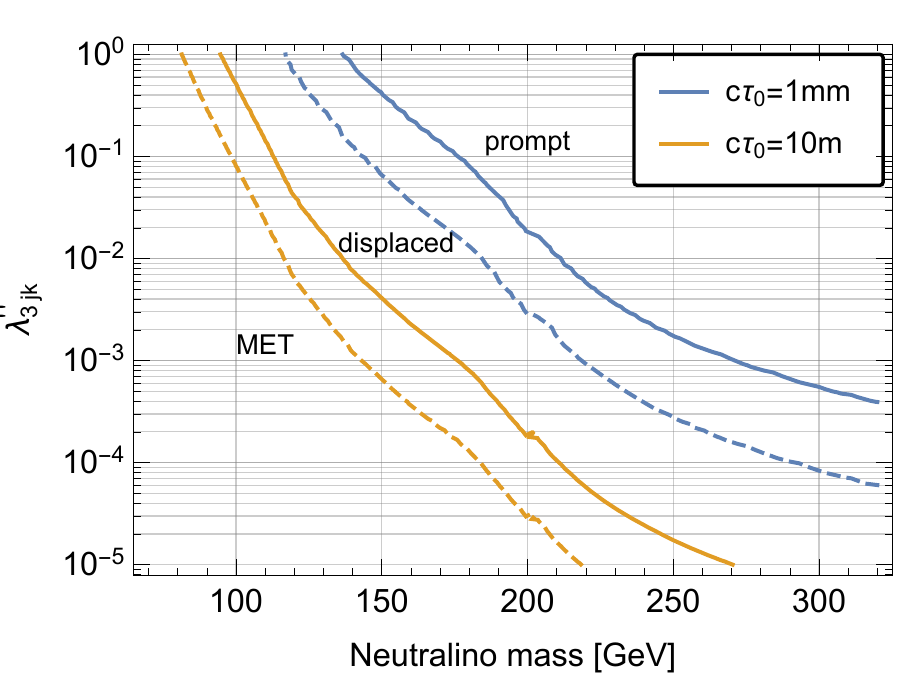}
\includegraphics[width=0.48\textwidth]{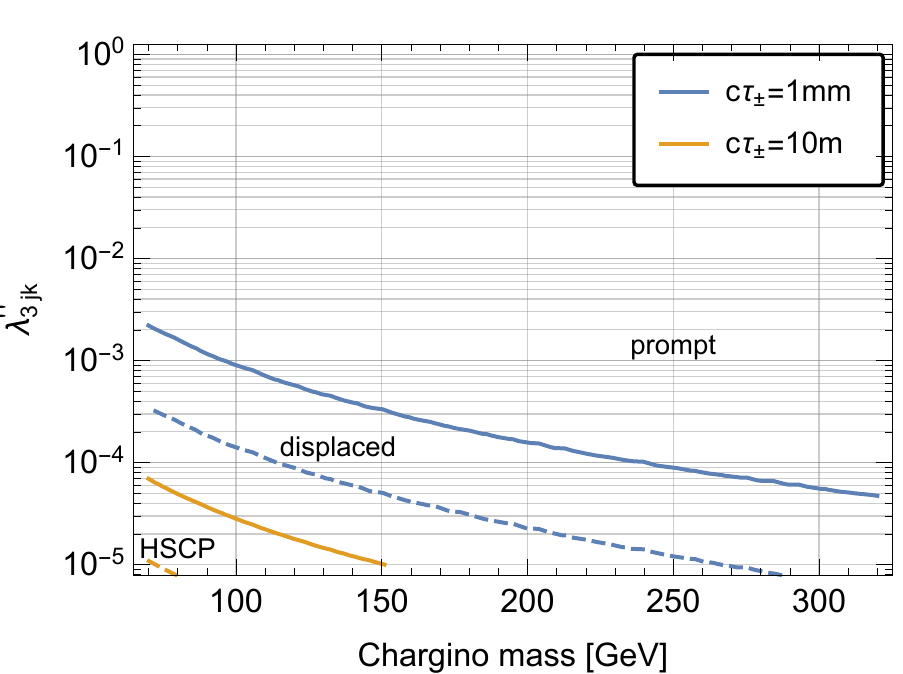}
\caption{Contours for the decay length of the neutralino, $\nino\to t d_j d_k$, (left) and of the chargino $\chinop\to \bar b \bar d_j \bar d_k$ (right), as a function of the RPV coupling $\ljk$ and the neutralino mass $\mtn$, with the stop mass set to 400\gev \ (dashed lines) and 1\tev\ (solid lines). The blue lines indicate displaced vertices with $c\tau=1$ mm, while the yellow lines indicate collider-stable particles, $c\tau=10$ m, giving either missing $E_T$ for the neutralino or heavy stable charged particles (HSCP) for the chargino.}
\label{fig:ninoctau}
\end{center}
\end{figure}

\subsection{Displaced vertices}\label{sec:dv}

So far we have not discussed weather the stop or neutralino decays are prompt or not. While the stop gives prompt decays over all the large $\ljk$ parameter space that we are considering, the same is not always true for a light neutralino: in particular, for $\mtn\ll m_t$, the neutralino decays via an off-shell top quark and can often give a displaced vertex or simply leave the detector as missing energy. In Fig. \ref{fig:ninoctau}, we set the stop mass to two values, 400\gev\ and 1\tev, and show contours of constant neutralino (left) and chargino (right) decay lengths. If they decay within the detector (1 mm $\lesssim c\tau\lesssim$10 m), displaced vertices searches will be pertinent, while for longer lifetimes the chargino is a heavy stable charged particle (HSCP) and the neutralino exits the detector as missing energy (MET). In this last case, for each stop the final state would be a top plus MET; if resonantly-produced this is a mono-top signature, while if the squark is pair-produced this is the standard scenario covered by \rpcing searches \cite{Aad:2015pfx,CMS-PAS-SUS-13-023,Khachatryan:2015pot}.

There are numerous analyses looking for displaced jets at the LHC \cite{Aad:2015rba,CMS:2014wda}; we will use result from independent groups that have re-casted the original searches to look for displaced RPV decays \cite{Cui:2014twa,Liu:2015bma,Csaki:2015uza,Zwane:2015bra}. Because of the low background, displaced decays are often more constrained than the corresponding prompt searches. For example, limits on pair-produced gluinos decaying to three jets are as high as 1.5\tev\ when $1\text{ mm}<c\tau<1$ m. In the present work, we will use the upper limits on the cross section of a pair-produced resonance decaying to three jets as a function of its mass and lifetime, as given in Ref. \cite[Fig. 7]{Cui:2014twa}. The presence of a top quark instead of a jet in our final state is not expected to change the results as the decay of the top gives back a $b$ jet. We cross-check the consistency of our result by considering pair-production of higgsinos and comparing the excluded region for a displaced higgsino to the limits found in Ref. \cite{Liu:2015bma}, and find good agreement with their results (see Appendix \ref{app:dv} for more details). 

The cross section for pair-produced stops is usually orders of magnitude higher than the cross section for pair-produced neutralinos, such that most neutralinos originate from  stop decays (unless the decay is very phase-space suppressed, $\mtt\ll \mtn+m_t$). We compute the cross section for electroweak pair-production of a neutralino with the package \texttt{Resummino 1.0.7} \cite{Debove:2009ia,Debove:2010kf,Debove:2011xj,Fuks:2012qx,Fuks:2013vua}: for a bino, this cross section also depends on the mass of first-generation squarks, so for definiteness we set $m_{\tq_{1,2}}=1\tev$ and $m_{\tq_{1,2}}=10\tev\gg \mtt$ as two benchmark points.\footnote{In the second case, bino electroweak production is negligible and neutralinos only come from stop decays.} For a higgsino, the cross section is independent of the squark masses and is much larger.
Finally, we compare the production cross section to upper limits on $\sigma(pp\to (jjj)\, (jjj))$ from Ref. \cite{Cui:2014twa}, which are of the order of $1-10$~fb in the range $c\tau\simeq1$cm$-1$m, and find excluded regions in the $\mtn-c\tau$ plane. These can be translated to limits in the $\mtt-\ljk$ plane, for a given neutralino mass. We defer to Appendix \ref{app:dv} for the detailed results and here just discuss the implications of displaced neutralino decays for stop production. We also do not explicitly discuss the case of a displaced neutralino decay arising from resonant stop production, $pp\to\st\to t\nino, \nino \stackrel{DV}\to (tjj)$: this would result in a prompt top and a single displaced vertex (a {\it quasi-mono-top}), while the recasted analyses consider two displaced vertices. Covering this signature would entail recasting the original ATLAS and CMS searches (which require only one displaced vertex) and we leave this to future work; nevertheless, we expect that this topology can be easily excluded.

\begin{figure}[t]
\begin{center}
\includegraphics[width=\textwidth]{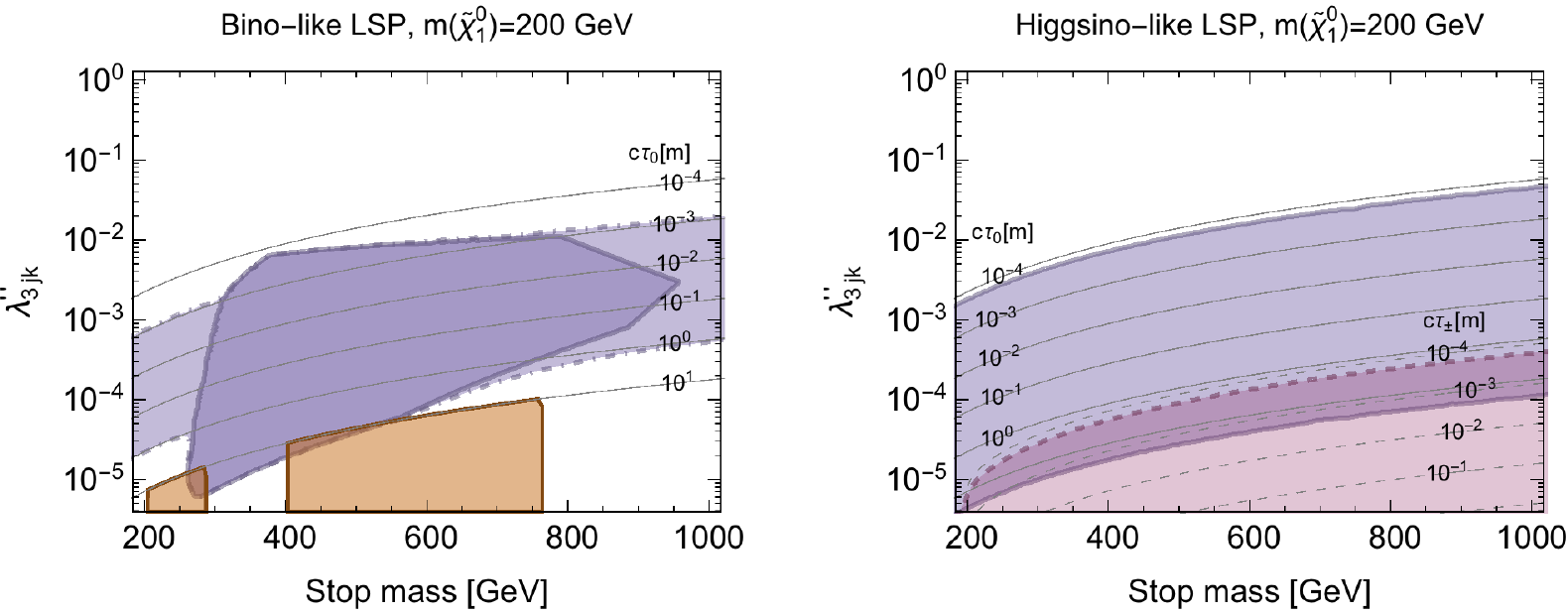}
\caption{Regions excluded by displaced decays and \rpcing searches. Left: in purple, regions excluded by displaced $\nino$ decay, for two choices of the mass of $\tilde q_{1,2}$,  $m_{\tq_{1,2}}=10\tev$ (decoupled squarks, solid line) and $m_{\tq_{1,2}}=1\tev$ (dot-dashed line). The orange excluded regions at the bottom have a collider-stable neutralino, $c\tau(\nino)>10$ m, corresponding to \rpcing searches. Right: for a higgsino-like LSP, the production cross sections are much larger, and a displaced decay is excluded for both the neutral (purple) and charged higgsinos (magenta). For $\mtn\simeq\mtch=200\gev$ the regions overlap. For reference, we also show contours of the neutralino and chargino lifetime.}
\label{fig:dvino}
\end{center}
\end{figure}

In Fig.~\ref{fig:dvino} we set the neutralino mass to 200\gev\ and display the excluded regions in the $\mtt-\ljk$ plane, for either a bino-like or a higgsino-like neutralino. Regions in purple are excluded by displaced decays of the neutralino to three jets, while the 
For the bino-like neutralino (left), we show the displaced decay limits for two choices of masses for first generation squarks, $m_{\tq_{1,2}}=1\tev$ and $m_{\tq_{1,2}}\gg \mtt$,  as regions respectively delimited by dot-dashed  and solid lines; for lower squark masses, first- and second-generation squarks would be abundantly pair-produced and give a same-sign top signatures via their RPV decay, and as such are excluded below $m_{\tq_{1,2}}\approx500\gev$ \cite{Durieux:2013uqa,Evans:2013jna}. Even for decoupled $\tq_{1,2}$, displaced neutralinos arising from stop decays give non-trivial constraints; as expected, those disappear for $\mtt\ll \mtn+m_t$, where the phase-space suppression of $\st\to t\nino$ limits bino production. Standard MET-based searches are shown in orange.
On the right panel, we show the exclusions for a higgsino-like neutralino: here the chargino and neutralino have the same mass and are abundantly produced, so that displaced higgsino decays are excluded in the whole lifetime ranges $0.5\times 10^{-3}\text{ m}<c\tau(\nino)<10$ m (purple region, see solid lines to read the neutralino lifetime) and $c\tau(\chinopm)>0.5\times 10^{-3}\text{ m}$ (magenta, see dashed lines for the chargino lifetime); these results agree with previous estimates \cite{Liu:2015bma,Csaki:2015uza}. In particular it should be noted that collider-stable charginos are excluded by HSCP searches up to approximately $700\gev$ \cite{Csaki:2015uza}, so that both a displaced and a long-lived chargino are excluded.

We have also considered the possibility of a collider-stable neutralino, $c\tau\gg10m$: here missing energy gives back some of the standard signature of the \rpcing MSSM: for a bino-like $\nino$, one can read  constraints from the searches for pair-produced stop followed by the decay $\st\to t\nino$ at high masses and  $\st\to c\nino$ or $\st \to b W^{*}\nino$ with an off-shell $W$ boson at low masses\cite{Aad:2015pfx,CMS-PAS-SUS-13-023,Khachatryan:2015pot}.\footnote{
In addition, a mono-top signature \cite{Andrea:2011ws,Boucheneb:2014wza} would arise from resonant stop production followed by the decay $\st\to t\nino$. While this is an interesting signature on its own, we find that the \rpcing searches at the LHC already exclude collider-stable neutralinos, up to $\mtt\simeq700\gev$. Monotop searches \cite{Aad:2014wza,Khachatryan:2014uma} might help raise the limits up to 1\tev, but we defer a precise assesment to future work.} Those limits are shown in Fig. \ref{fig:dvino} as light orange regions at the bottom.
For a collider-stable higgsino-like $\nino$, there is no corresponding \rpcing signature, because the decay to the chargino suppresses $\st\to t\nino$ and the subsequent RPV decay of the chargino $\chinop\to b d_j d_k$ dominates over RPC processes such as the phase-space suppressed decay $\chinop\to \nino W^{(*)}$. Depending on the chargino lifetime there are different signatures: the chargino decay length can be written as (see also Fig.~\ref{fig:ninoctau})
\beq
c\tau (\chinop\to \bar b \bar d_j \bar d_k)\simeq 0.8\text{ mm} \frud{10^{-3}}{\ljk}^2 \frd{100\gev}{\mtch}^5 \fru{\mtt}{1\tev}^4\,
\eeq
If the chargino decays promptly, it forms a three-jet resonance: in this case, the most relevant searches are the ones for pair-produced gluinos decaying into three jets \cite{ATLAS:2012dp,Chatrchyan:2013gia}. While the limits on the cross section are orders of magnitude larger than the charged higgsino cross section, charginos are also produced via stop decays: then, the relevant cross section is $\sigma(pp\to \st\st^*)\times Br(\st\to \chinop b)^2$, which unfortunately is still a factor of a few smaller than the experimental upper limits. If the chargino decay is displaced or long-lived, it is excluded by displaced and HSCP searches discussed above.

For $\mtn\simeq\mtch\gtrsim200\gev$, the range in which the neutral higgsino is collider-stable always corresponds to a displaced or long-lived chargino, so that any non-prompt higgsino is excluded \cite{Csaki:2015uza} (compare the higgsino excluded regions in Fig.~\ref{fig:dvino}, where the excluded regions overlap and extend downward, to Fig.~\ref{fig:dv_hino} in the Appendix, which shows that for different neutralino masses a collider-stable neutralino is still allowed).

To summarize, displaced neutralino decays give non-trivial constraints. A combination of displaced, \rpcing and HSCP searches excludes most of parameter space with $c\tau(\nino)\gtrsim 10^{-4}$ m, only leaving slivers of parameter space such as $\mtn<\mtt<\mtn+m_t$ for a bino-like LSP. A displaced higgsino is excluded in the whole range $10^{-4}$ m$\lesssim c\tau(\nino)\lesssim10$ m, $10^{-4}$ m$\lesssim c\tau(\chinopm)$.

\subsection{Combined limits with a light neutralino}
\begin{figure}[t!]
\begin{center}
\begin{tabular}{|l|c|c|c| }\hline
Signature & Label & Color & Experimental Searches \\\hline
$pp\to \st \to jj $ & \texttt{DIJET} & \swatch{MMA1} &\cite{Aad:2014aqa,Khachatryan:2015sja,CMS-PAS-EXO-14-005,Khachatryan:2015dcf}\\
$pp\to \st\st^* \to (jj)(jj)$ & \texttt{PAIRED DIJETS} & \swatch{MMA3} & \cite{Khachatryan:2014lpa,ATLASCONF2015026}\\
$pp\to \st \to ttX$, $pp\to \st^*\to\bar t\bar t X$ & \texttt{SSTOP} & \swatch{MMA2} &\cite{Chatrchyan:2013fea}\\
$pp \to \st\st^*\to  tt\bar t\bar t X$ &\texttt{4TOP}& \swatch{MMA4} & \cite{Chatrchyan:2013fea,Aad:2015kqa}
\\
$pp\to \st\st^* \to (t\nino)(\bar t\nino), pp\to \tilde \chi^{0,\pm}\chi^{0,\pm}$ & \multirow{2}{*}{\texttt{DVjjj}} & \multirow{2}{*}{\swatch{MMA5} }&\multirow{2}{*}{ \cite{Cui:2014twa,Liu:2015bma,Csaki:2015uza}}\\
\hfill with displaced $\tilde\chi^{0,\pm}$ decays & &&\\\hline
\end{tabular}
\end{center}
\begin{center}
\captionsetup[subfigure]{labelformat=empty,position=top}
\subfloat[\quad Bino-like LSP, $\mtn=200\gev$]{
\includegraphics[width=0.47\textwidth]{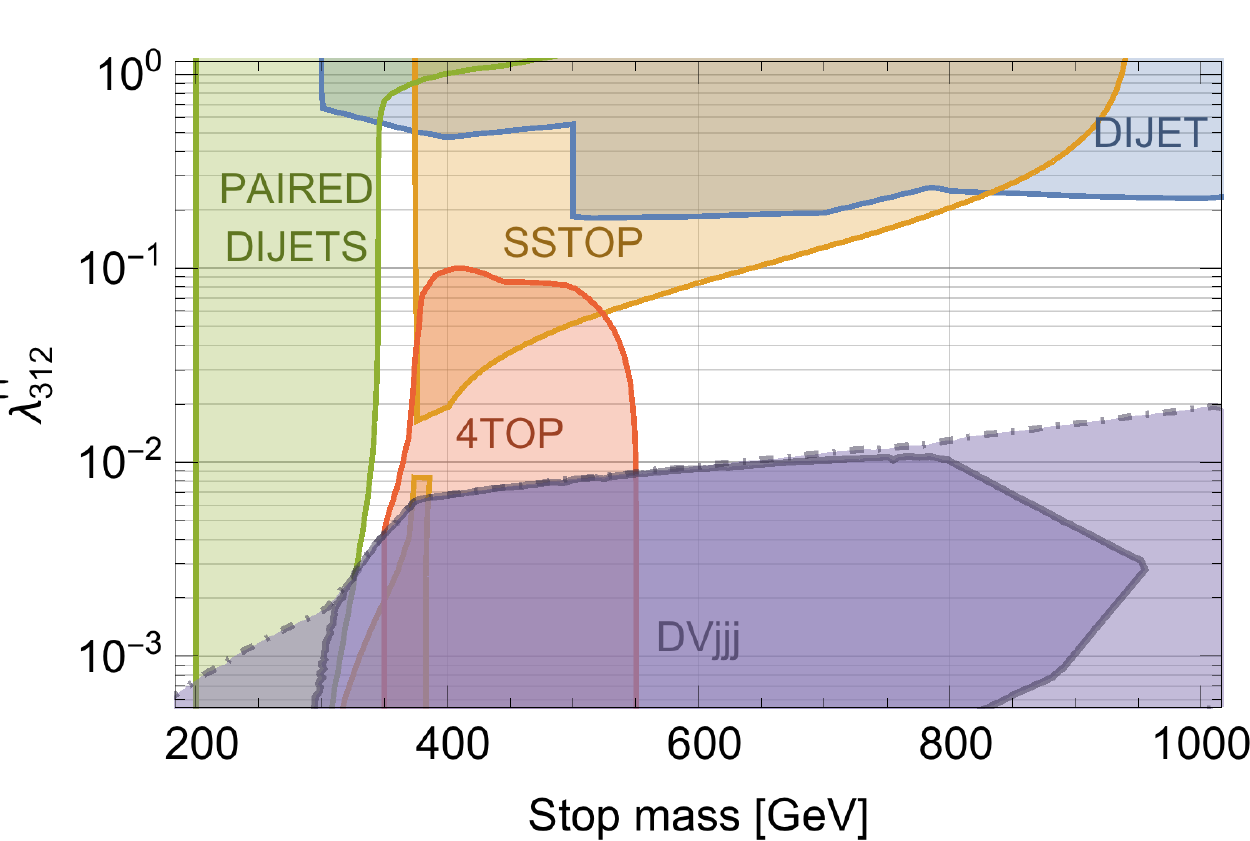}
}\hfill
\subfloat[\qquad Higgsino-like LSP, $\mtn=200\gev$]{\includegraphics[width=0.47\textwidth]{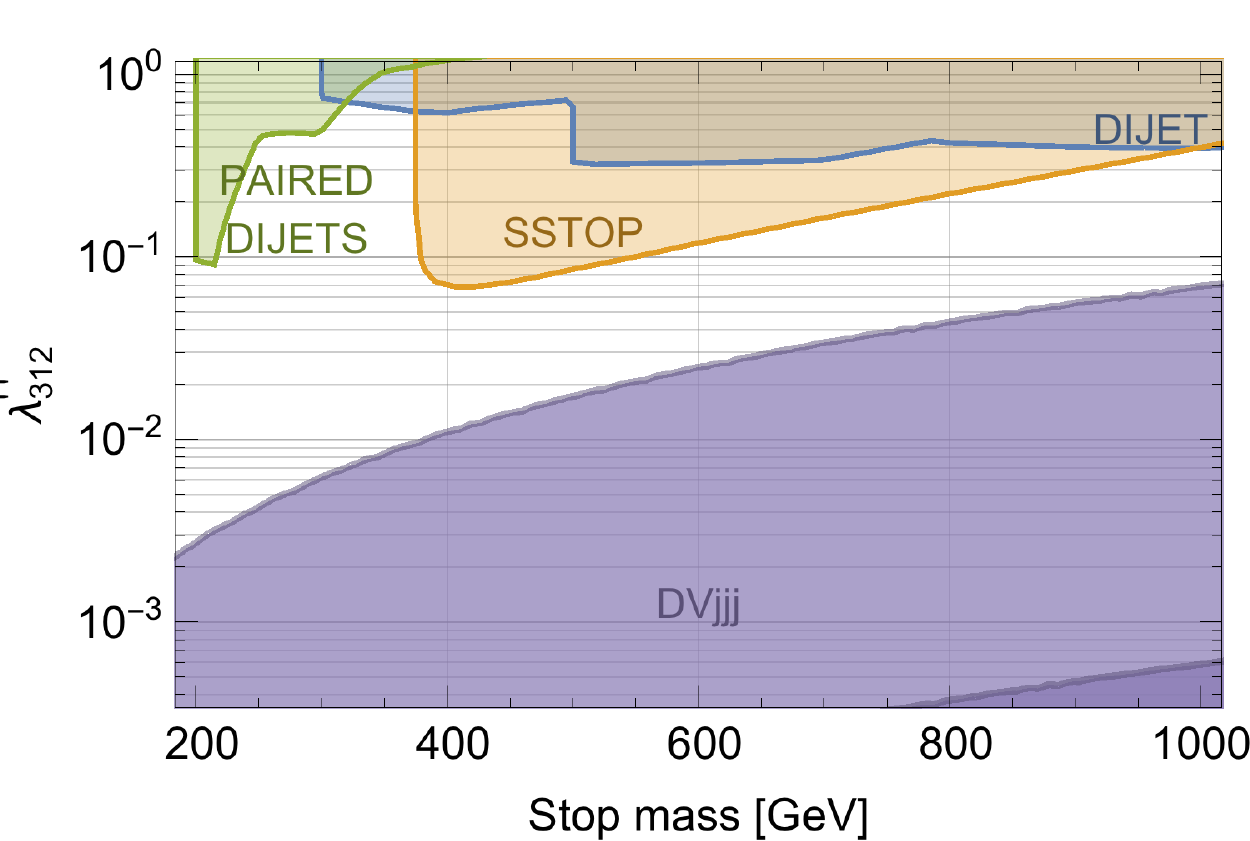}
}
\\
\includegraphics[width=0.47\textwidth]{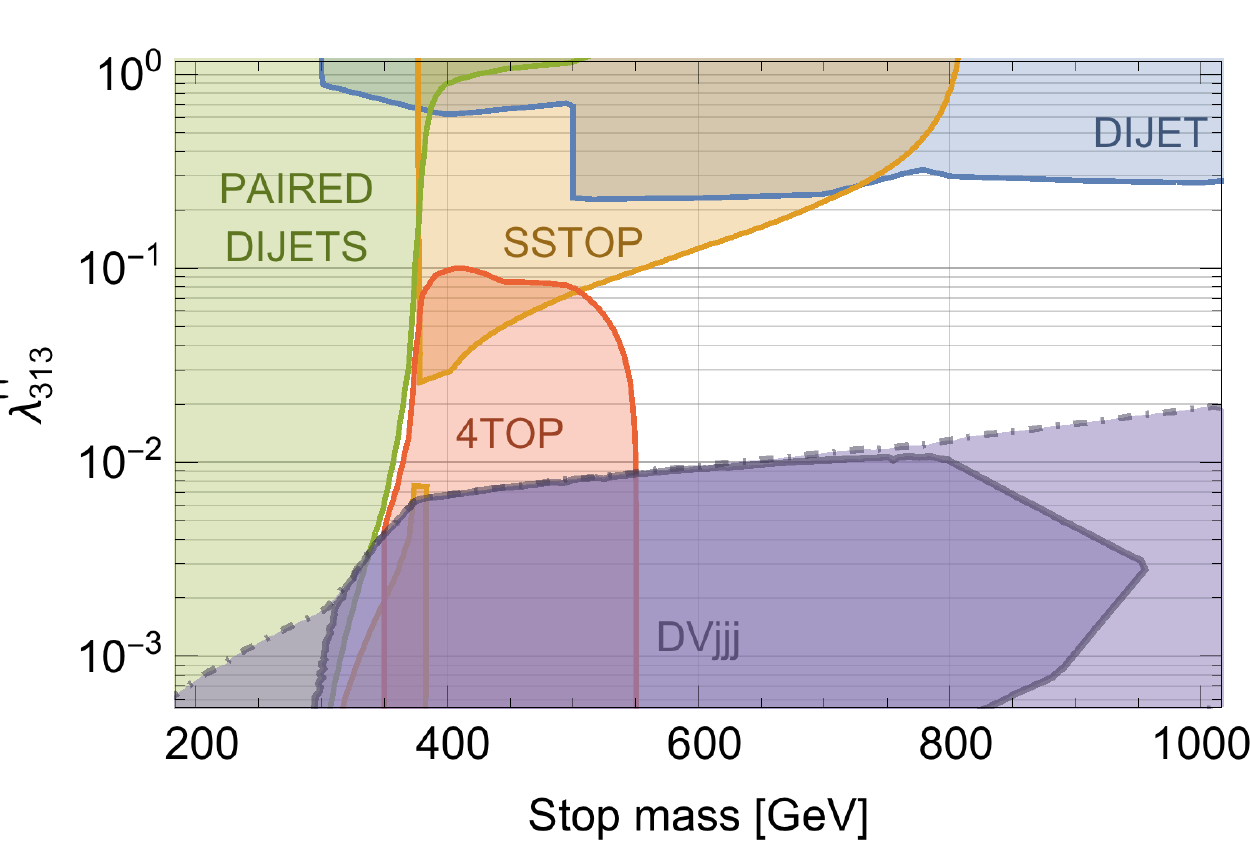}\qquad
\includegraphics[width=0.47\textwidth]{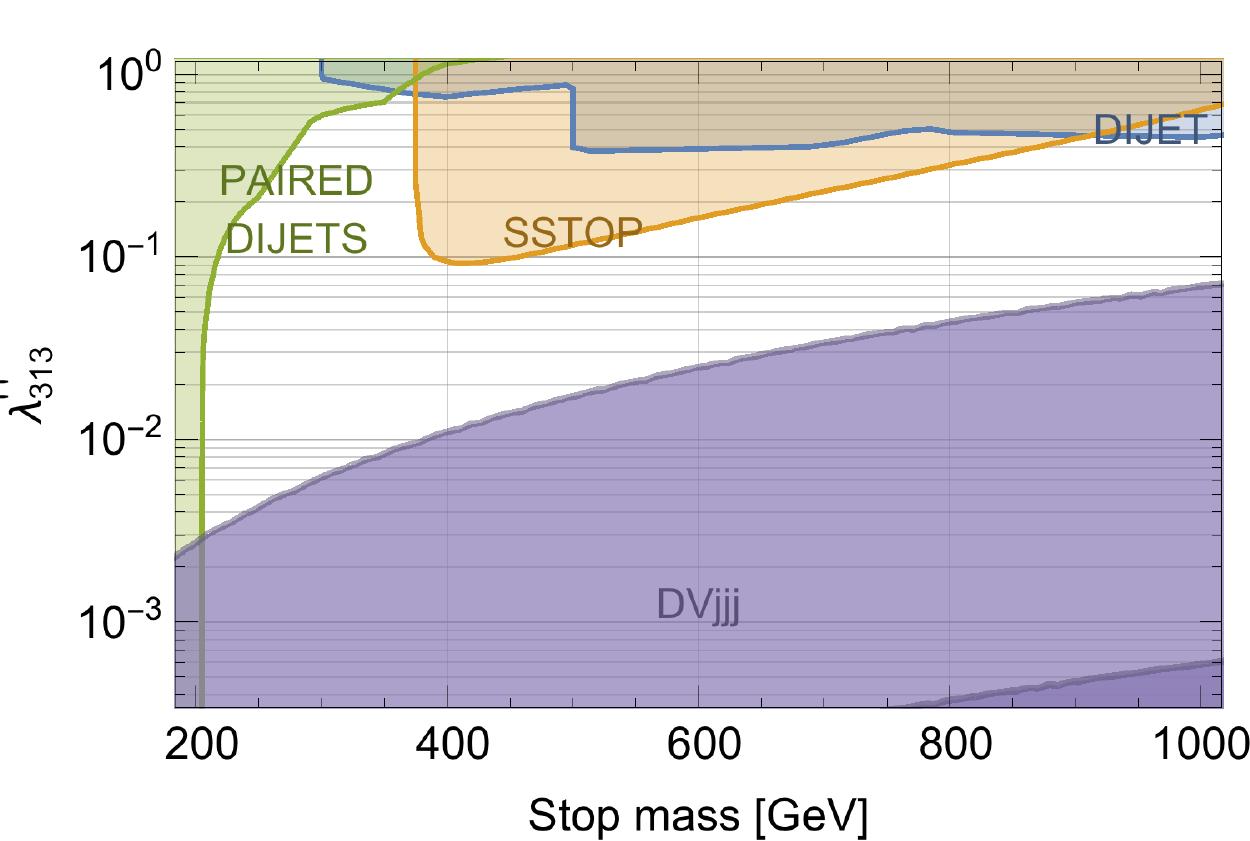}\\
\includegraphics[width=0.47\textwidth]{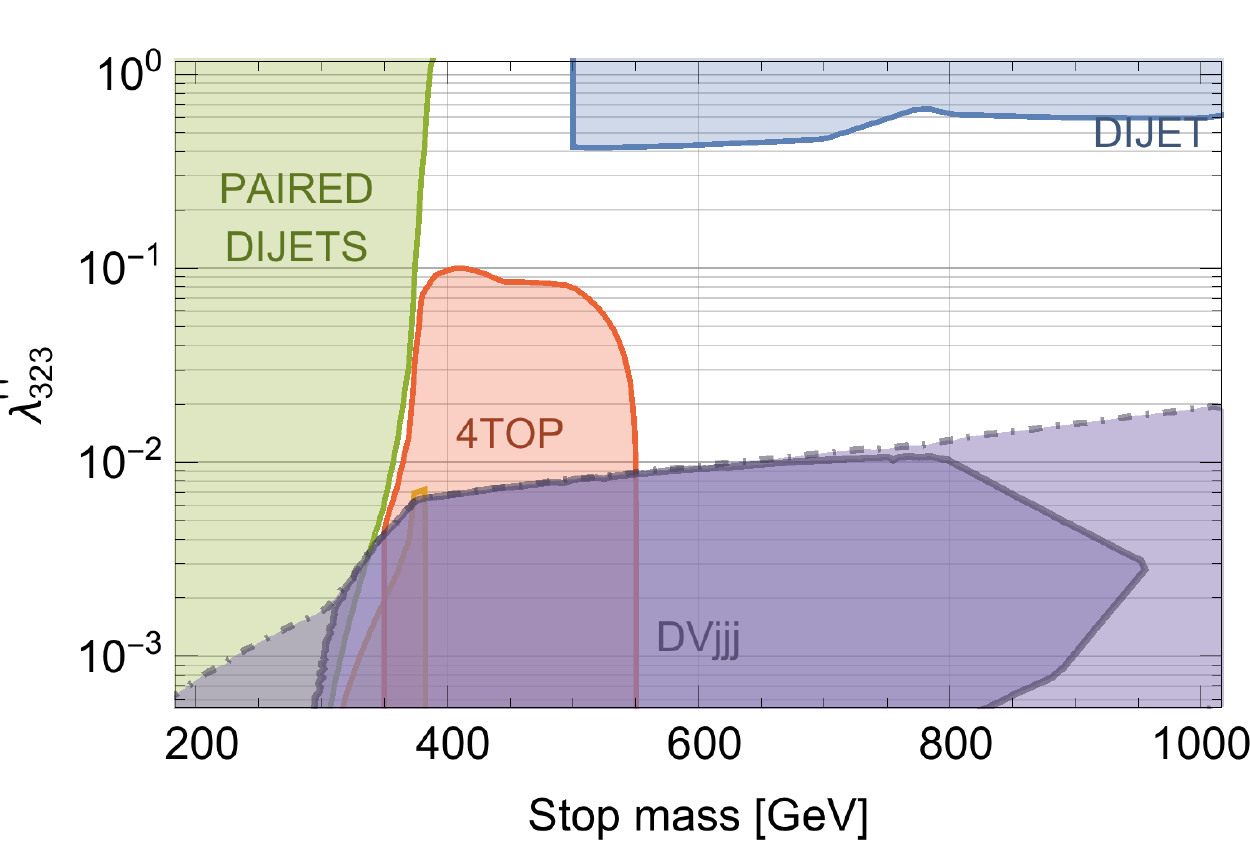}\qquad
\includegraphics[width=0.47\textwidth]{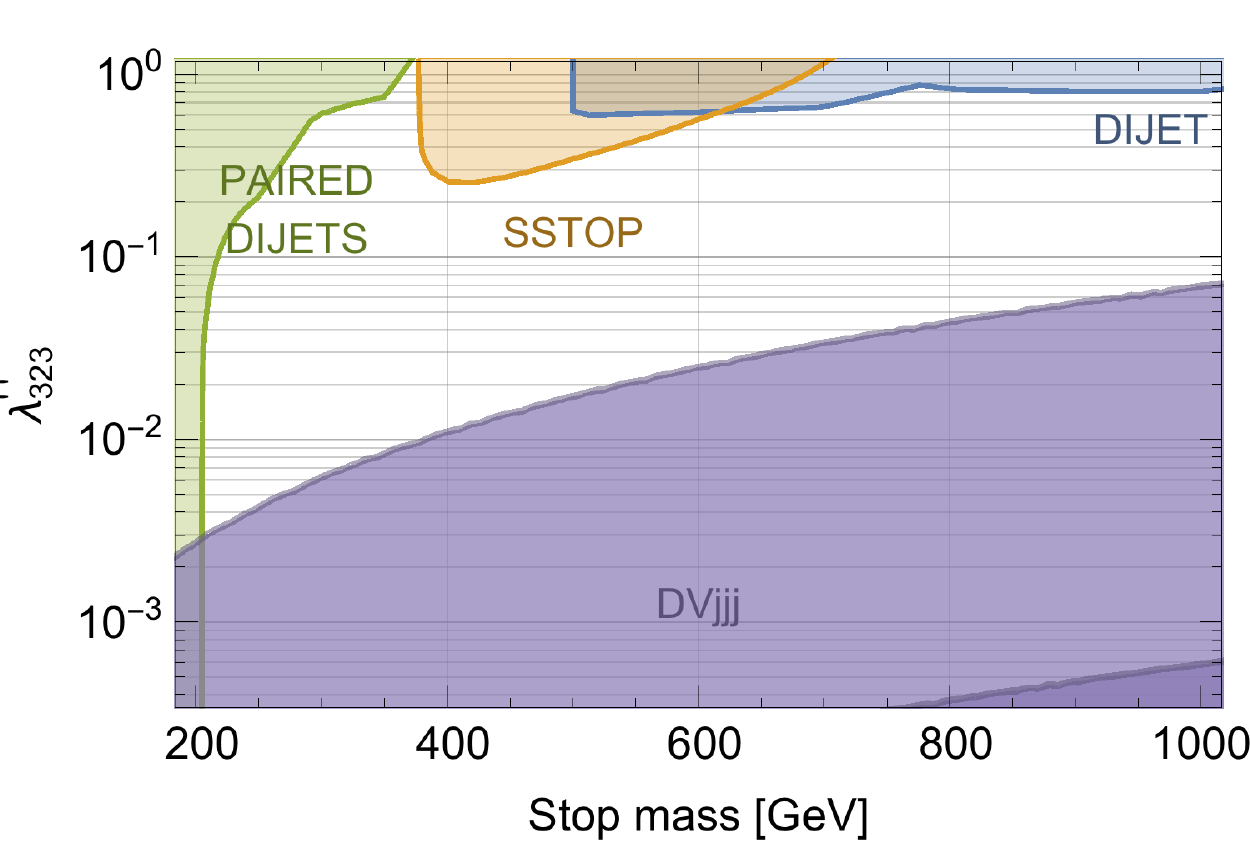}
\caption{Combined exclusion limits for each RPV coupling $\l_{312}$ (top row), $\l_{313}$ (middle row) and $\l_{323}$ (bottom row), for a bino-like (left column) and for a higgsino-like neutralino LSP with $\mtn=200\gev$. Each colored region is excluded by experimental searches as detailed in the table at the top.
}
\label{fig:combined_ino}
\end{center}
\end{figure}

We can now summarize all the previous signatures and show combined limits in the $\mtt-\ljk$ plane in Fig. \ref{fig:combined_ino}, where the neutralino mass is set at 200\gev. The regions excluded by  the relevant ATLAS and CMS searches are color-coded according to their experimental signatures, as detailed at the top of the figure.

Limits from resonant production cover regions with large coupling, while signatures of pair-produced stops cover small couplings. It can be seen how the limits  based on dijet resonances (respectively, same-sign tops) become weaker at small (large) $\ljk$. Some stop masses are definitely excluded: for example, stops between 400 and 500 \gev \ are forbidden for $\l_{312},\l_{313}$ if the LSP is bino-like. For clarity, we have not extended the vertical range downward to include the possibility of long-lived neutralinos resulting in \rpcing signatures of pair-produced stops \cite{Aad:2015pfx,CMS-PAS-SUS-13-023,Khachatryan:2015pot}. The corresponding excluded region for the bino can be read off of Fig. \ref{fig:dvino}. Note that in the higgsino case the displaced vertices exclusions for $\nino$ and $\chinopm$ overlap and the purple region sets {\it lower limits} on the \rpving couplings.

\begin{figure}[!h]
\begin{center}
%
\includegraphics[width=0.4\textwidth]{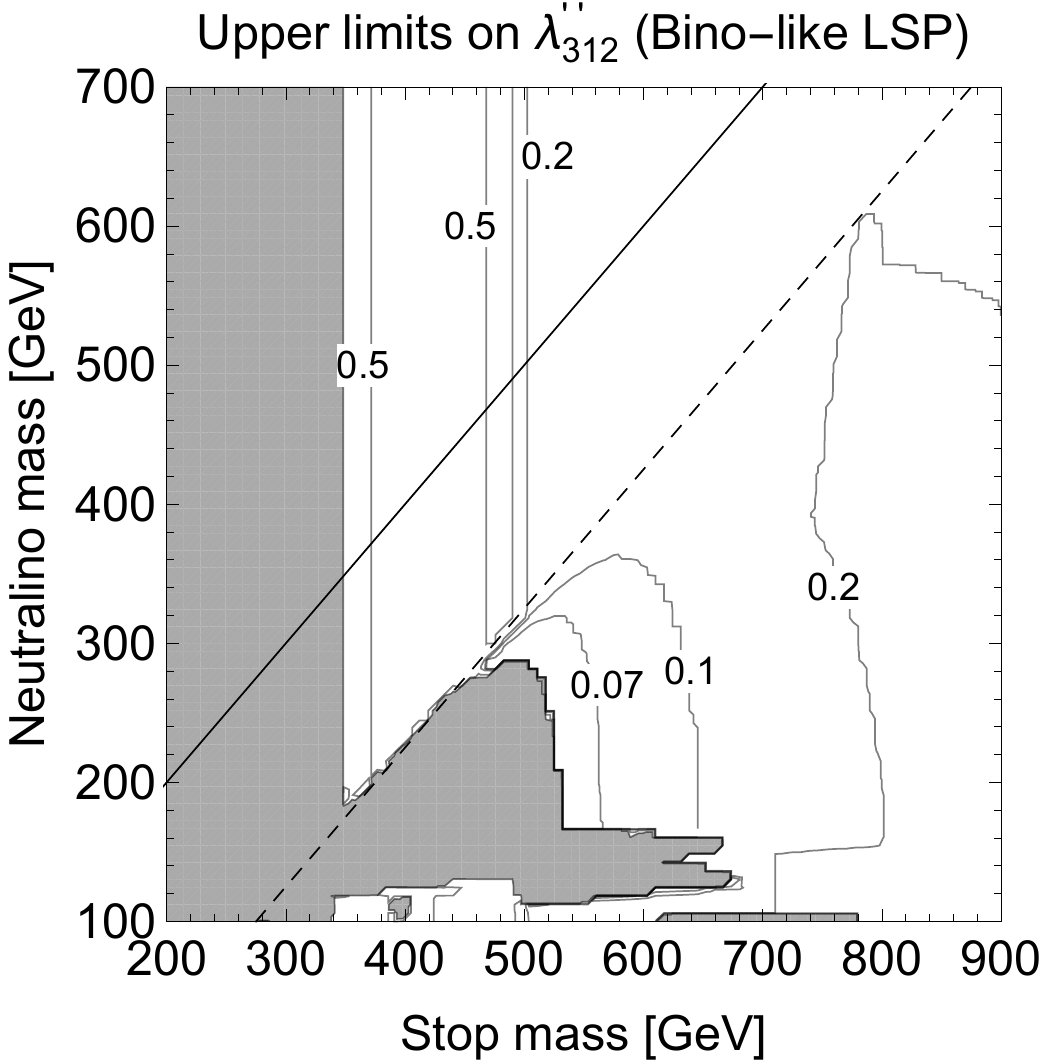}\qquad\qquad
\includegraphics[width=0.4\textwidth]{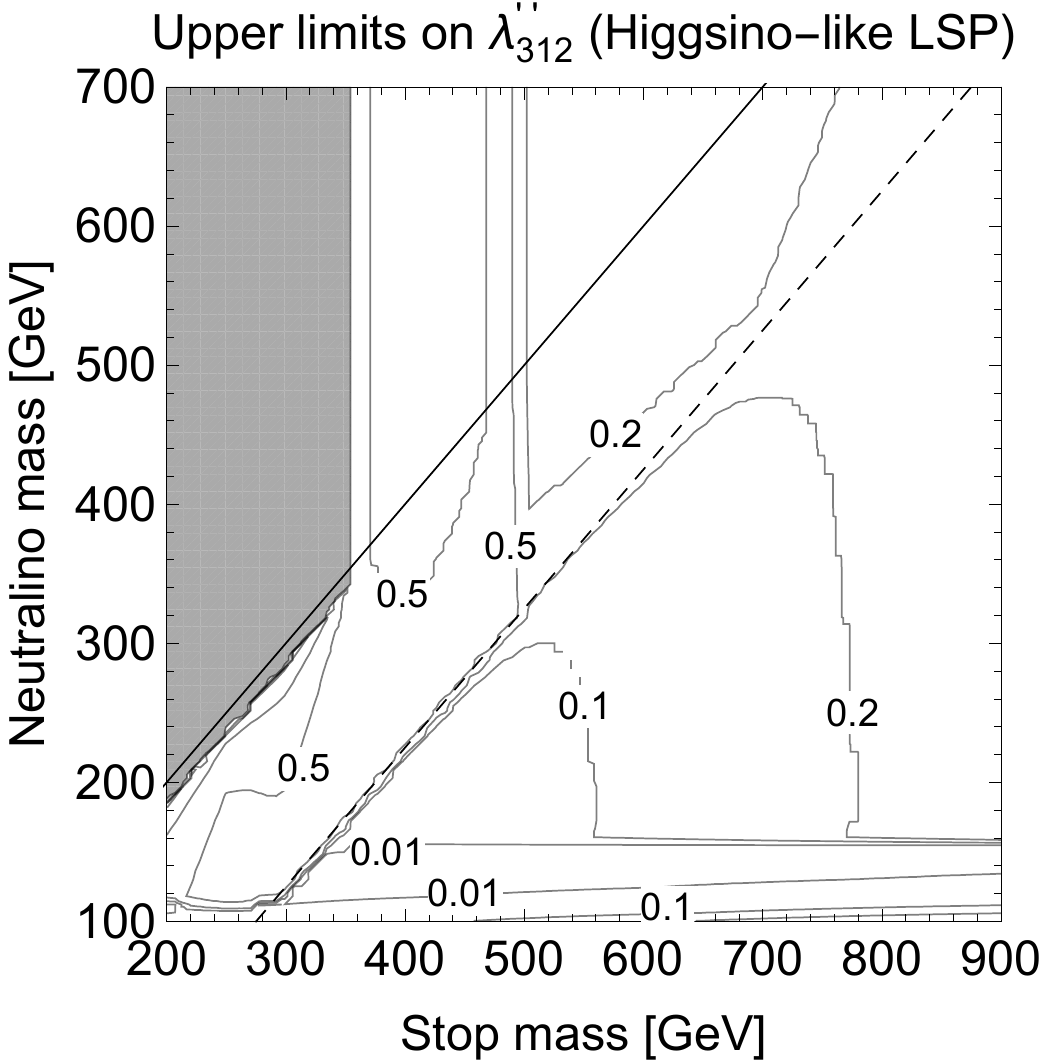}

\medskip

\includegraphics[width=0.4\textwidth]{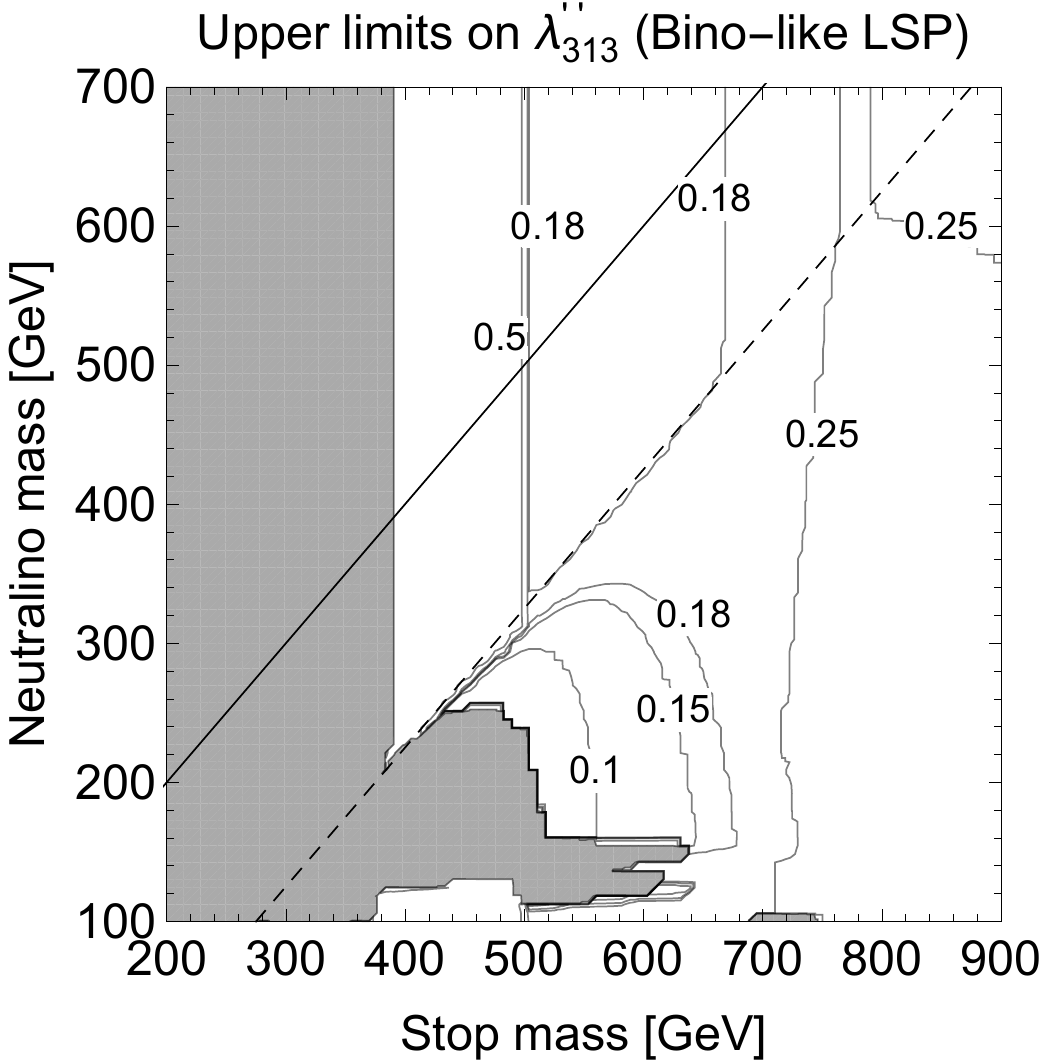}\qquad\qquad
\includegraphics[width=0.4\textwidth]{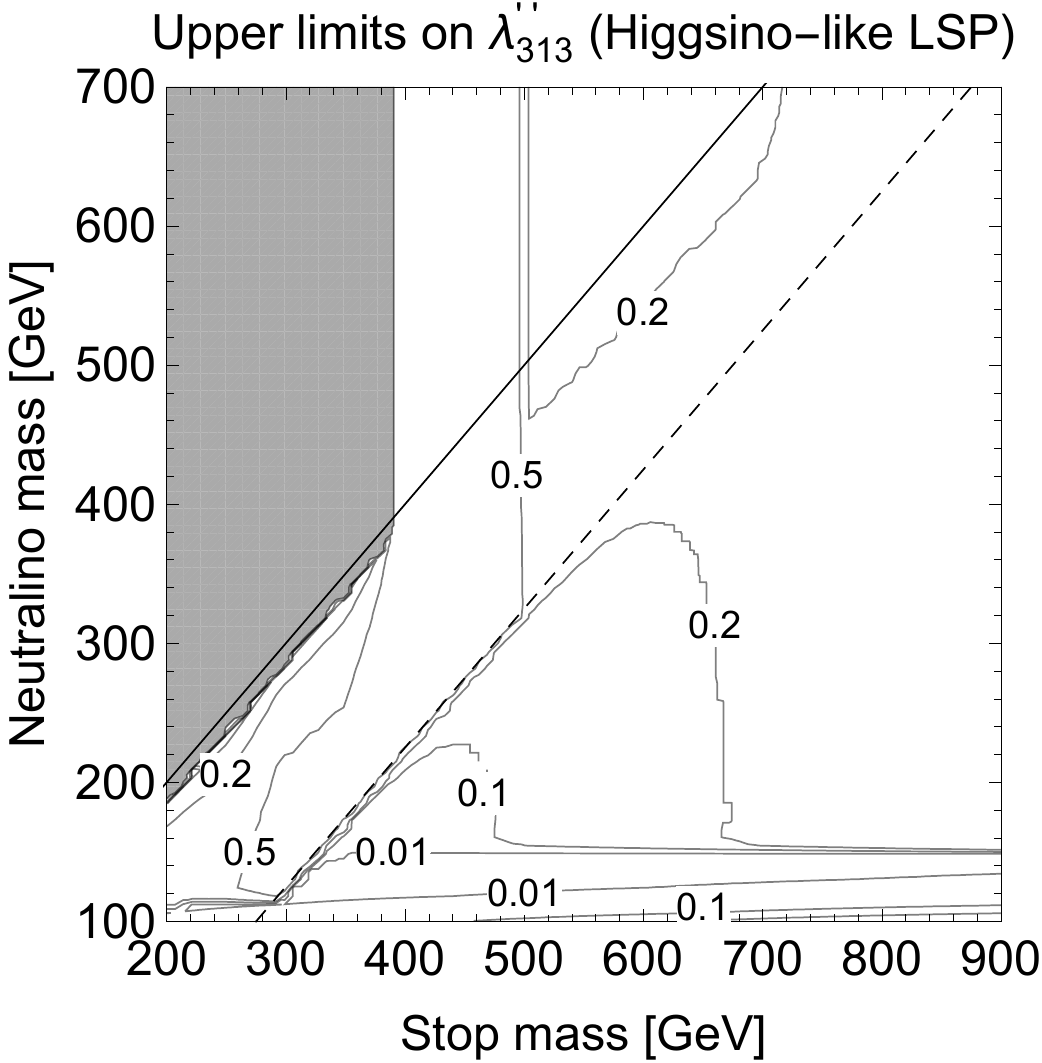}

\medskip

\includegraphics[width=0.4\textwidth]{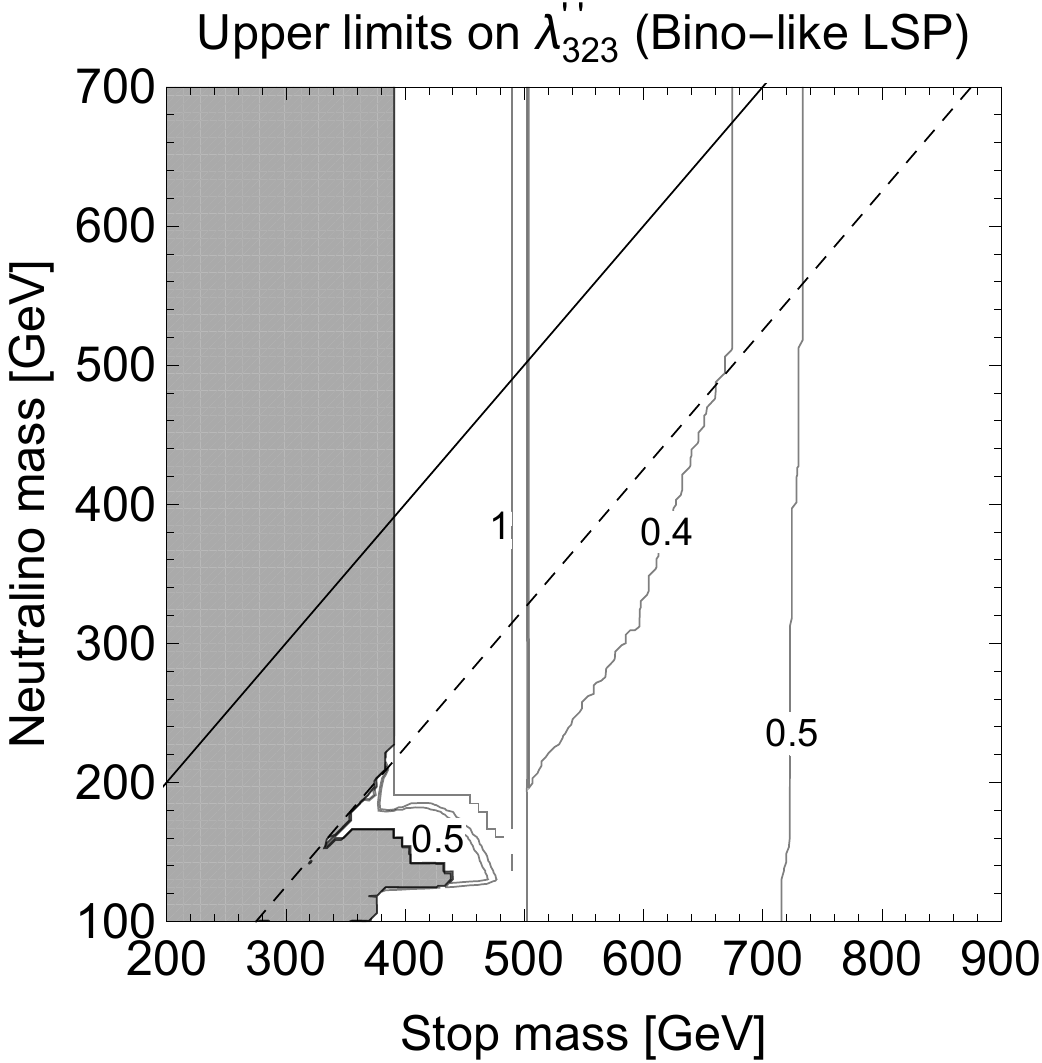}\qquad\qquad
\includegraphics[width=0.4\textwidth]{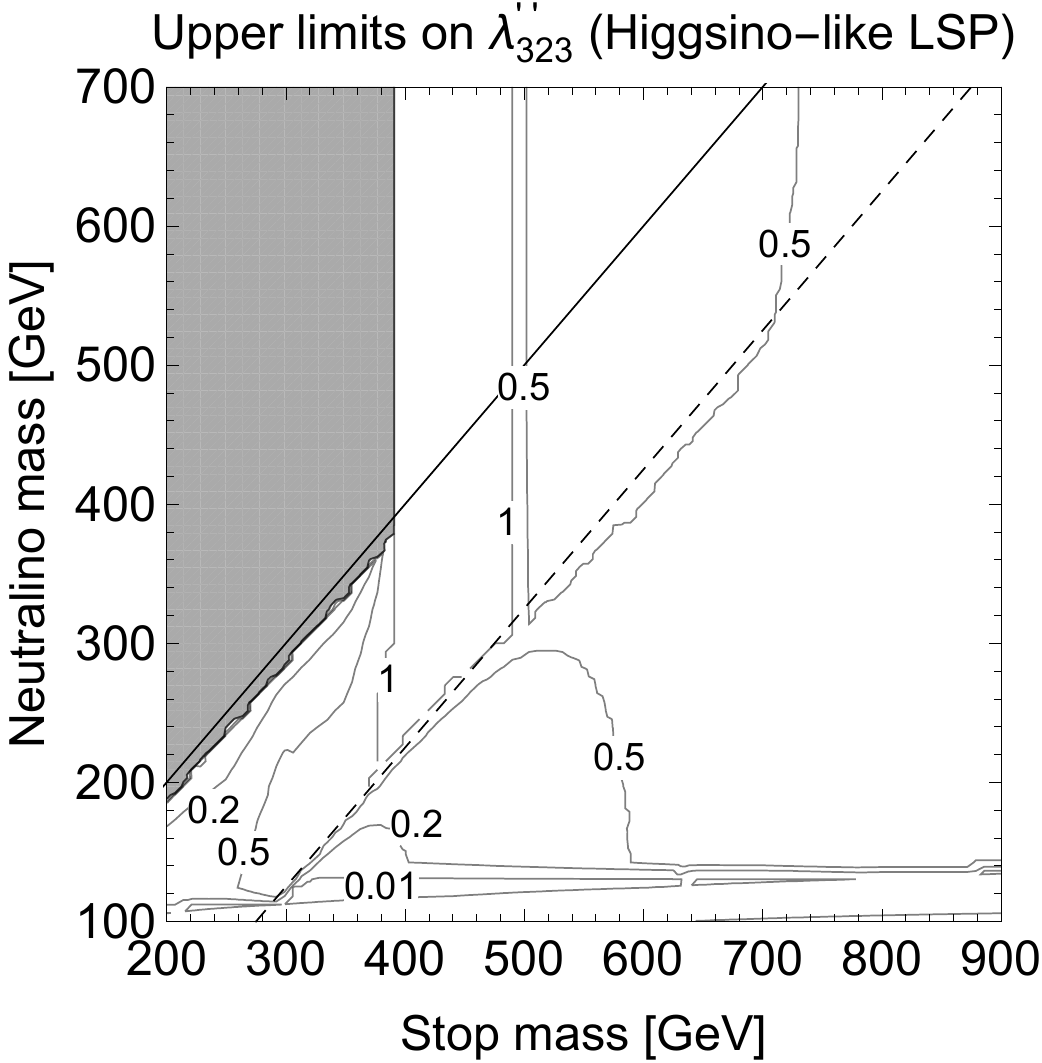}
\caption{Upper limits on each RPV coupling $\l_{312}$ (top row), $\l_{313}$ (middle row) and $\l_{323}$ (bottom row), for a neutralino LSP that is bino-like (left column) or higgsino-like (right column), as a function of the stop and neutralino masses. The gray regions are excluded {\it for all values} of $\ljk$. The diagonal lines correspond to $\mtt=\mtn$ (solid) and $\mtt=\mtn+m_t$ (dashed). Above the diagonal lines, the results in Sec.~\ref{sec:stopLSP} for a stop LSP apply, while below them we have the signatures discussed in this section. 
}
\label{fig:combined_contours}
\end{center}
\end{figure}

\afterpage{\clearpage}


Those result hold for a specific choice of the neutralino mass. To show the explicit dependence on $\mtn$, in Fig.~\ref{fig:combined_contours} we show the upper limits on each \rpving coupling as a function of the stop and neutralino masses. Gray regions are excluded by combining all the searches discussed.
Qualitatively, there are three regions of parameter space: first, for a stop LSP (above the solid diagonal line) the dijet constraints of Sec.~\ref{sec:stopLSP} hold, independently from the neutralino mass;  for $\mtn< \mtt<\mtn+m_t$ (above the dashed diagonal line), the same limits apply for a bino-like neutralino LSP, as the stop decay $\st\to t\nino$ is suppressed, while weak constraints on the higgsino-like LSP are given by single and pair-produced dijet resonances; third, for $\mtt> \mtn+m_t$ (below the dashed diagonal line) limits from same-sign tops give the strongest limits, with constraints from displaced neutralino decays for $\mtn\lesssim m_t$. Light bino-like neutralinos at $\mtn\simeq100\gev$ are not completely excluded as the resonant SS top channel loses sensitivity when the neutralino is long-lived (although we expect that a single displaced neutralino decay would also be excluded). We can draw the following conclusions from Fig. \ref{fig:combined_contours}:

\begin{itemize}
\item With a bino-like neutralino, large \rpving couplings, $\ljk\simeq1$, are excluded  for almost  any sub-TeV stop, apart for $\l_{323}$ in the range $\mtt\approx 400-500\gev$. This small region is still viable due to the mild excess in the same-sign dilepton analysis \cite{Chatrchyan:2013fea} and the absence of strong limits on dijet resonances below 500\gev. A right-handed stop can be excluded up to 500\gev\ if decays are prompt and up to 700\gev\ from displaced neutralino decays (again, apart from the small region around $400-500\gev$ with $\l_{323}>0.1$). The exclusions from dijet resonances go well above 1\tev, see Fig. \ref{fig:stopLSP} for the multi-TeV region.
\item For a higgsino-like neutralino, \rpv\ with O(1) couplings is excluded for right-handed stops below 1\tev. At the same time, the range $\ljk\lesssim 0.1$ {\it cannot be probed with present data} and specific searches targeting the chargino decay $\chinopm\to b d_j d_k$ are needed. For a pair-produced stop, this decay mode was studied in Ref. \cite{Evans:2013uwa} where the authors re-casted the ATLAS multijet analysis \cite{Aad:2015lea} and found that the signal cross section was just below the upper limits, thus giving no constraints. For a single-produced stop, there are no searches specifically targeting a four-jet resonance; we discuss this signature in Sec. \ref{sec:new}.
\item For large neutralino masses, stop decays such as $\st\to t\nino$ do not give appreciable limits in the region $\mtt\ll\mtn+m_t$, where they are phase-space suppressed. This is similar to the \rpcing case, and decays such as $\st\to b W\nino$ and $\st\to c\nino$ will become more important. Thus, there are weak limits in between the paired dijet and the multi-top exclusions. At the same time, the neutralino lifetime is reduced and  displaced vertices appear only at very small $\ljk$.
\item For smaller neutralino masses, displaced vertices are possible at large $\ljk$: limits on pairs of displaced resonances are strong and we can exclude stop masses up to 700\gev\ for $\mtn\simeq 150\gev$ in the bino case. 
One can notice a blind spot near $\mtt=400\gev,\ \mtn=100\gev$: here neutralinos are displaced/collider-stable and the prompt multi-top signatures disappear. We expect this region to be effectively probed by the displaced jet resonance searches, which only required one displaced vertex. For the higgsino case, we can exclude large ($\gtrsim 10^{-2}$) \rpving couplings for $\mtn<150\gev$\ and stop masses below 1\tev\, as in this region higgsinos would  decay within the detector. 
\end{itemize}

\section{Prospects for LHC13 and proposed new searches}\label{sec:lhc13}
While Figs.~\ref{fig:combined_ino}-\ref{fig:combined_contours} already give strong limits on stops with large \rpv, there are signatures that are currently not being examined by ATLAS or CMS. For example, a higgsino LSP could still be as light as 100\gev, with a stop at 200\gev\ and \rpving couplings larger than 0.1! To probe this range, we propose new searches that target the chargino decay channel of the stop, as well as an optimized same-sign top search that will be more sensitive in other regions. In addition, collisions at 13 \tev\ give larger cross sections for all of the signals discussed, and we will show the projected reach of the upgraded LHC run.

\subsection{New searches}\label{sec:new}

\paragraph{4j resonances} A {\it natural} neutralino sector generically has light charginos and new searches should be performed to investigate the chargino-mediated stop decay
\beq
\st \to  b \chinop \to  b \bar b \bar d_j \bar d_k\,.
\eeq
This correspond to a {\it four-jet resonance}, with two or three $b$-jets (one of the final $d_j$'s might also be a $b$ quark) and a {\it three-jet sub-resonance} within the primary resonance. The natural width of the chargino resonance is always much smaller than the detector mass resolution. Depending on the production channel, different strategies would be optimal:
\begin{itemize}
\item {\it single four-jet resonances}: in this case, four jets arise from the resonant production of the stop. Typically, there is one $b$-jet with large transverse momentum, $p_T(b_1)=O(\mtt-\mtn)$, and three softer jets from the chargino. All jets are spatially well separated, $\Delta R\gtrsim 1$, and the three non-leading jets form a narrow three-jet resonance. Loosening the requirements for the CMS three-jet paired resonance search \cite{Chatrchyan:2013gia} ($N_{\text{jet}}\geq4$ instead of $N_{\text{jet}}\geq6$ ) could already identify the chargino. The simultaneous reconstruction of  a three and a four-jet resonance would be a striking signature of this model. 

The $H_T$ distributions have an edge at $H_T\simeq \mtt$ and rapidly fall off at larger values: at low stop masses it could be hard to trigger on a signal with $H_T<300\gev$, but at the same time the potentially large cross section can make it feasible to use the tail of the distribution. To discrimante against the background, one can also profit from multiple $b$-tags and the different $p_T$ distributions of the various jets and impose harder cuts on the $b$-jet with highest $p_T$. In Fig. \ref{fig:pT} we show the parton-level $p_T$ distributions of the jets from a 300\gev\ stop via a 200\gev\ chargino, before and after the following cuts:
$$
\begin{array}{|c|cccc|c|}\hline
\multirow{2}{*}{$H_T^{min}$ [\text{GeV}]} & \multicolumn{4}{c|}{p_T^{min} \ [\text{GeV}]}& \multirow{2}{*}{$\epsilon_{cuts}$} \\\cline{2-5}
 & b_1 & b_2 &j_1 &j_2&\\\hline
 350 & 80 & 30 & 100 & 50& 0.0019\\\hline
\end{array}
$$
In the figure we display the expected number of events with an integrated luminosity of 20\,\ifb \ at 13\tev, given production through  $\l_{312}=0.1$ and taking into account an efficiency of O(0.7) for each $b$-tagged jet. The jets are sorted by transverse momentum, $p_T(b_1)>p_T(b_2)$ and $p_T(j_1)>p_T(j_2)$ (this discussion can be generalized in case of production via $\l_{313},\l_{323}$ which gives three $b$-jets). In this case, the low acceptance can be balanced by the large cross section, $\sigma\simeq 50$ pb, resulting in a large number of events, $O(1000)$ for $L=20\,\ifb$. Such a search could cover the current blind spot with a higgsino LSP and low stop masses in Fig.~\ref{fig:combined_contours}.
\begin{figure}[t]
\begin{center}
\includegraphics[width=\textwidth]{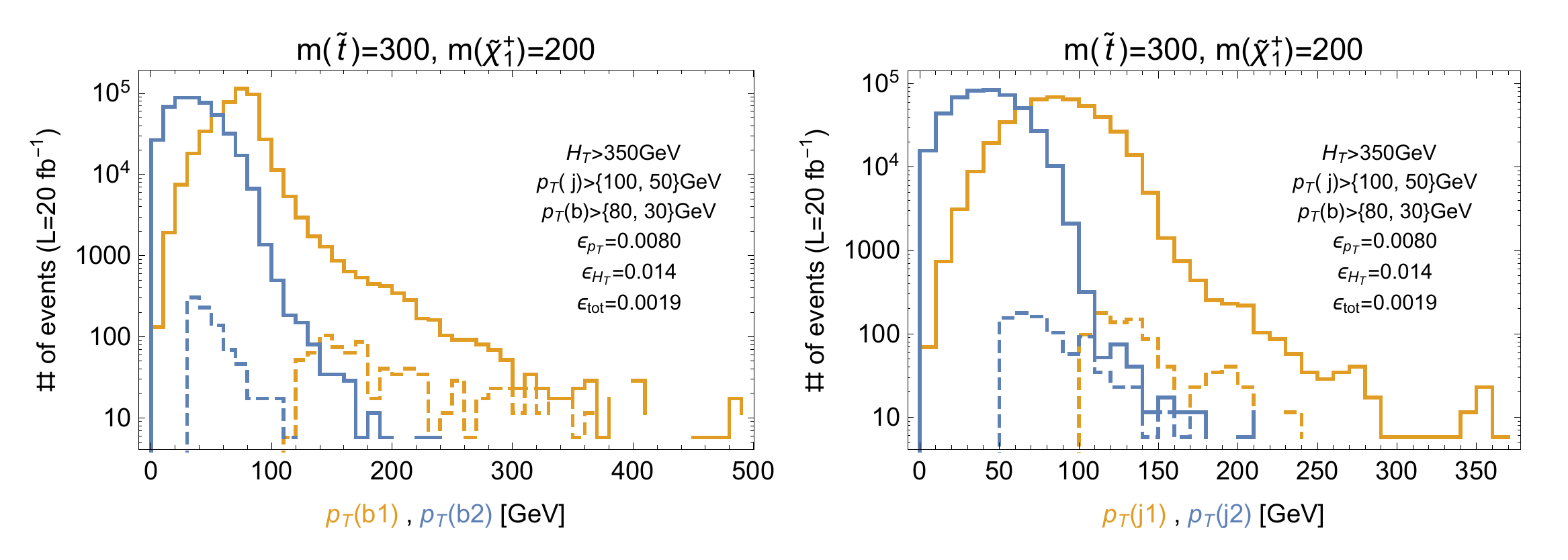}
\caption{Parton-level $p_T$ distributions of the four jets originating from the resonant production of a stop followed by the decay $\st\to \bar b \chinop \to \bar b b j j $ (with $\mtt=300\gev$, $\mtc=200\gev$ and $\sqrt s=$13 TeV). Imposing the listed cuts brings the solid lines down to the dashed lines.}
\label{fig:pT}
\end{center}
\end{figure}

At larger stop masses and/or larger gaps between stop and chargino masses, the signal is much cleaner and can be easily triggered. Given that the cross section is proportional to the RPV coupling and that the signal yield does not decouple at large $\ljk$ (see Eq~\eqref{eq:nondecoupling}), this signature has the potential of excluding squarks with large RPV couplings up to the multi-TeV range  (see black long-dashed line in Fig.~\ref{fig:lhc13}), which could have direct implications for many scenarios of baryogenesis.

\item{\it pair-produced four-jet resonances}: here, the search strategy would be similar to the CMS RPV gluino search \cite{Chatrchyan:2013gia}, except for the stricter requirement $N_{\text{jets}}\geq8$ instead of $N_{\text{jets}}\geq6$. Ref.~\cite{Evans:2014gfa} showed that 8\tev \ data could already bring limits on the stop between 600 and 700\gev. Requiring a three-jet resonance within the four-jet resonance would increase the rejection against the background. Additionally, even an unchanged three-jet search itself can independently probe the chargino three-jet decay: the 13\tev\ cross section for stop pair-production followed by chargino decay will increase above current (8\tev) gluino limits for $\mtt\lesssim 400\gev$ making it feasible to indirectly exclude stops via their decay products if the 13\tev \ limits are not much higher. Other options for this signal are cut-and-count analyses (similar to the ATLAS search \cite{Aad:2015lea}), with at least eight jets and requiring four or six $b$-tagged jets: the authors of Ref.~\cite{Evans:2013uwa} recasted the ATLAS 8\tev\ analysis and found that limits on the cross sections were a factor of a few weaker than the production cross section.  At 13\tev, the increased cross section will make it possible to probe this channel.

\end{itemize}
While this searches explicitly targets the chargino decay mode of the stop, they would also apply to first- and second-generation squarks, resonantly produced  through $\l_{2jk}$ couplings or pair-produced as usual: here, both the neutralino and the chargino decay modes of the squarks are prompt and result in multi-jet resonances.

\paragraph{SStop+2j} The usual multi-top searches \cite{Chatrchyan:2013fea,Aad:2015kqa} only require same-sign leptons and the presence of jets from the top quark decay, ($N_{\text{jets}}\geq2, N_{\text{b-jets}}\geq1,2$ for Ref.~\cite{Chatrchyan:2013fea}); defining search regions with two additional jets for each top quark pair (one of which could be a $b$-jet) would give better constraints on our model. While the additional jets are soft for $\mtn\simeq m_t$, they would be more easily identified for heavier neutralinos ($p_T(j_1)\simeq p_T(j_2)= O(\mtn-m_t)$).

Within this decay topology, exotic final states such as $ttt\bar t$ or $ t \bar t \bar t \bar t$ (+4j) are also possible when stops are pair-produced: those are currently not investigated in multi-top searches \cite{Aad:2015kqa}, but they hold great potential.

\subsection{LHC13}
With Run 2 of the LHC just started, we can see how much more constrained the large RPV scenario can become in the immediate future. The higher parton luminosities increase the cross sections by factors of about  4 and 14 for resonant and pair-production of stops at 1\tev, but the increase can be much larger for resonant stop production in the multi-TeV range, e.g. $\sigma_{\text{13\tev}}/\sigma_{\text{8\tev}}\sim50-100$ for a 4 TeV stop.

We do not perform a full detector simulation, mainly because the experimental cuts that will be used by ATLAS and CMS are yet not public (in addition, multiple 13\tev\ searches will be based on new scouting techniques, which will be able to reach much lower masses than current searches); the limits will also depend on the integrated luminosity at the time of each data release (see Ref.~\cite{Duggan:2013yna} for a sensitivity study of some signatures of pair-produced squarks with 300 and 3000 \ifb). Instead of giving expected limits for a given luminosity (as is usually done in phenomenological studies), a useful exercise is to see what happens if, in the absence of any discovery, the experimental upper limits on the cross sections for each corresponding signature reaches a certain value. For definiteness, we take these ``projected'' limits to be the current  limits based on 20\ifb \ at 8 TeV; this naturally takes into account the weakening of the limits at lower masses, where the bakgrounds are larger. This will not be achieved at a specific value of the integrated luminosity, as signal and background scale differently with the increased energy (for example, for a given luminosity of $20\,\ifb$ the exclusions will be weaker at lower masses given the larger background, while at high masses one can expect a similar reach), but we can estimate that they will be reached with O(20-50) \ifb.\footnote{This estimate is based on the expectation that the larger backgrounds in the sub-TeV range will be somewhat balanced by the use of scouting techniques.} With this in mind, the increase in the production cross sections from larger parton luminosities will be enough to improve the current limits on the RPV couplings.

We show the prospective improvements in Fig. \ref{fig:lhc13}, where we keep the same color-coding as in Fig. \ref{fig:combined_ino}: limits from Fig.~\ref{fig:combined_ino} are demarked by colored dashed lines and the expected excluded regions are enclosed by solid lines with matching color for each signature. In the table above the figure, we list the values of the ``projected limits'' on the cross sections on which the exclusions are based.  As soon as 13\tev\ experimental limits are made public, the reader can look at the actual limit on the production cross section for a certain signature (and at a given mass) and compare it to the hypothetical limit listed in the top table. If the actual limits are stronger than those listed, the excluded regions will be larger. Otherwise, the excluded regions will be smaller, in between the solid and dashed lines of the same color.

 In addition to the reach of the same searches of Fig. \ref{fig:combined_ino}, in grayed-out regions we also show the prospective reach of searches targeting the chargino decay $\chinopm\to bb d_j d_k$, also detailed in Sec. \ref{sec:new}. 
The {\bf black, long-dashed line} gives the expected reach of a new search for a resonantly produced stop followed by decays to a chargino, $\st\to b\chinop\to b\, \bar b\bar d_j \bar d_k$ (four-jet resonance); here a simple ansatz was that the expected upper limits would be similar to the limits on pair-produced dijet resonances, although we expect that a dedicated analyses could do better.
The {\bf black, medium-dashed line} gives the expected excluded region for the pair-produced stop decaying into four jets, where we have kept the experimental upper limits of the re-casted multijet analysis in Ref. \cite{Evans:2013uwa}. Finally, the  {\bf black, short-dashed line} shows the expected exclusion region from the three-jet resonance searches looking for only the charginos coming from a pair-produced stop. In this case, the ansatz for the ``expected'' experimental limits was taken as the exclusions of the CMS gluino search for pair-produced three-jet resonances \cite{Chatrchyan:2013gia}, discriminating whether the final state includes light or heavy flavors.

\afterpage{%
\begin{figure}[t!]
\thispagestyle{empty}
\vspace{-5mm}
\begin{center}
\begin{tabular}{|l|ccccc|c|c|}\hline
&\multicolumn{5}{c|}{Stop mass [GeV]}& Line/ & \multirow{3}{*}{\rotatebox[origin=c]{270}{Searches}}\\
\cline{2-6}
Process & 200 &400 & 600 & 800 &1000 & filling&\\
&\multicolumn{5}{c|}{LHC13 ``projected'' upper limits  [pb]}&style& \\\hline
$pp\to \st \to jj $ & -  & 160 & 2.4 & 1.45 & 0.39 & \coline{MMA1}& \multirow{5}{*}{\rotatebox[origin=c]{270}{Existing}}\\
$pp\to \st \to tt+X $ & 0.035 & 0.035 & 0.035 & 0.035 & 0.035  & \coline{MMA2}&\\
$pp\to \st\st^* \to (jj)(jj)$ & 4.05 & 0.52 & 0.17 & 0.13 & 0.037 & \coline{MMA3}&\\
$pp\to \st\st^* \to (bj)(bj)$ & 5.7 & 0.44 & 0.14 & 0.086 & 0.092 & \coline{MMA3}&\\
$pp \to \st\st^*\to  tt\bar t\bar t+ X$ & - & 0.049 & 0.0183 & 0.0085 & 0.0041 & \coline{MMA4}&\\
\hline
$pp\to \st \to (bbjj)$ & 4.05 & 0.52 & 0.17 & 0.13 & 0.037 & \dashLong{black}&\multirow{3}{*}{\rotatebox[origin=c]{270}{New}}\\
$pp\to \st\st^* \to (bbjj)(bbjj)$&- & 2.67 & 0.45 & 0.12 & 0.064 & \dashlong{black}&\\
$pp\to \st\st^* \to (bjj)(bjj)+X$& 16.7 & 0.46 & 0.24 & 0.081 & 0.062 & \dash{black}&\\\hline
\end{tabular}
\captionsetup[subfigure]{labelformat=empty,position=top,margin=0pt,parskip=0pt}
\subfloat[\quad Bino-like LSP, $\mtn=200\gev$\vspace{-1mm}]{
\includegraphics[width=0.47\textwidth]{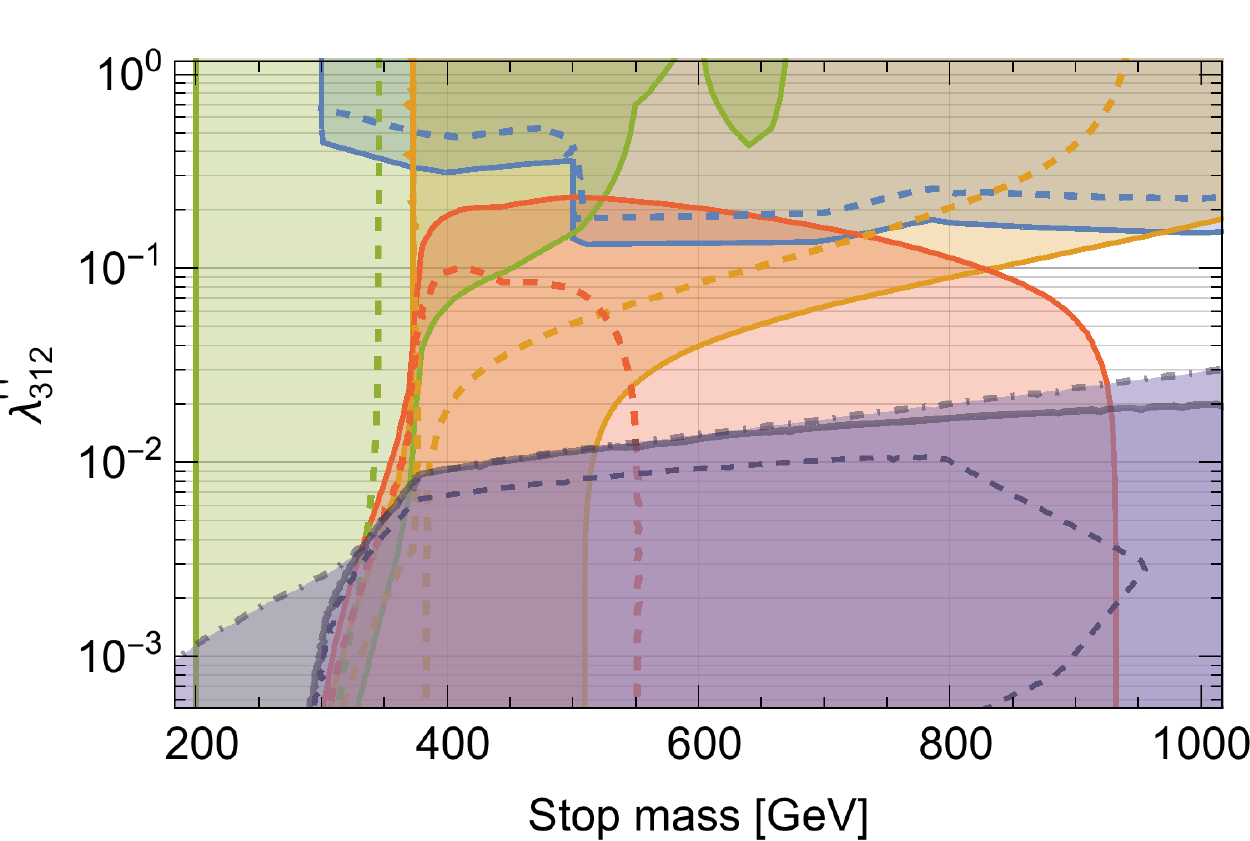}
}\hfill
\subfloat[\qquad Higgsino-like LSP, $\mtn=200\gev$\vspace{-1mm}]{\includegraphics[width=0.47\textwidth]{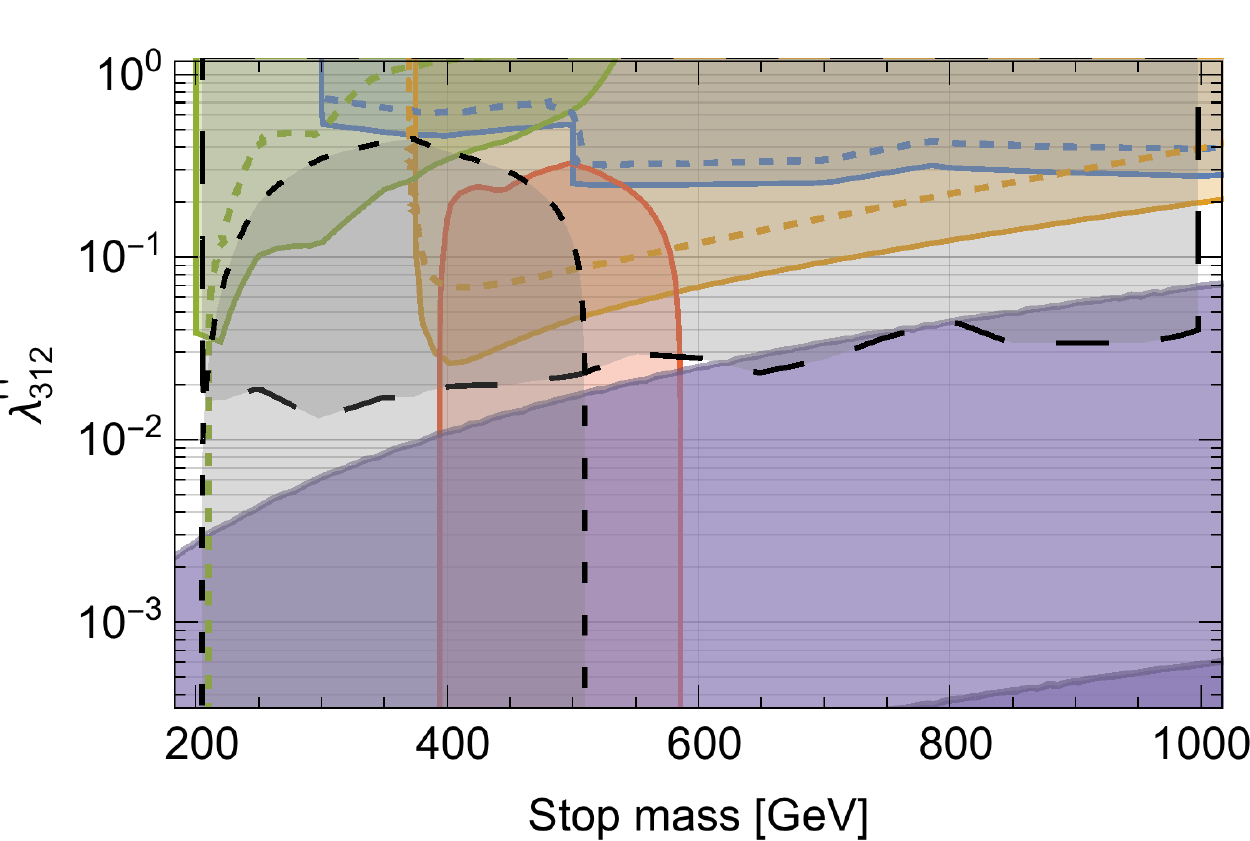}
}
\\
\includegraphics[width=0.47\textwidth]{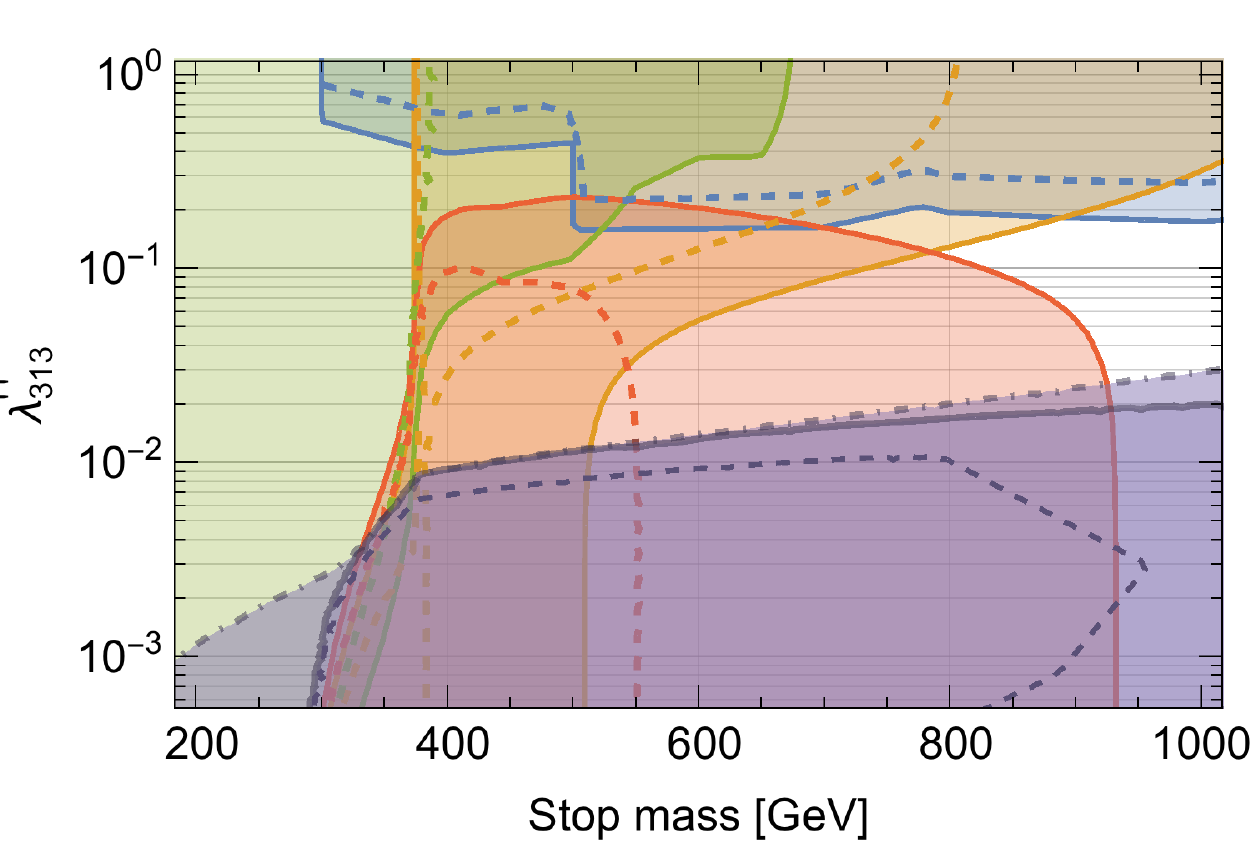}\qquad
\includegraphics[width=0.47\textwidth]{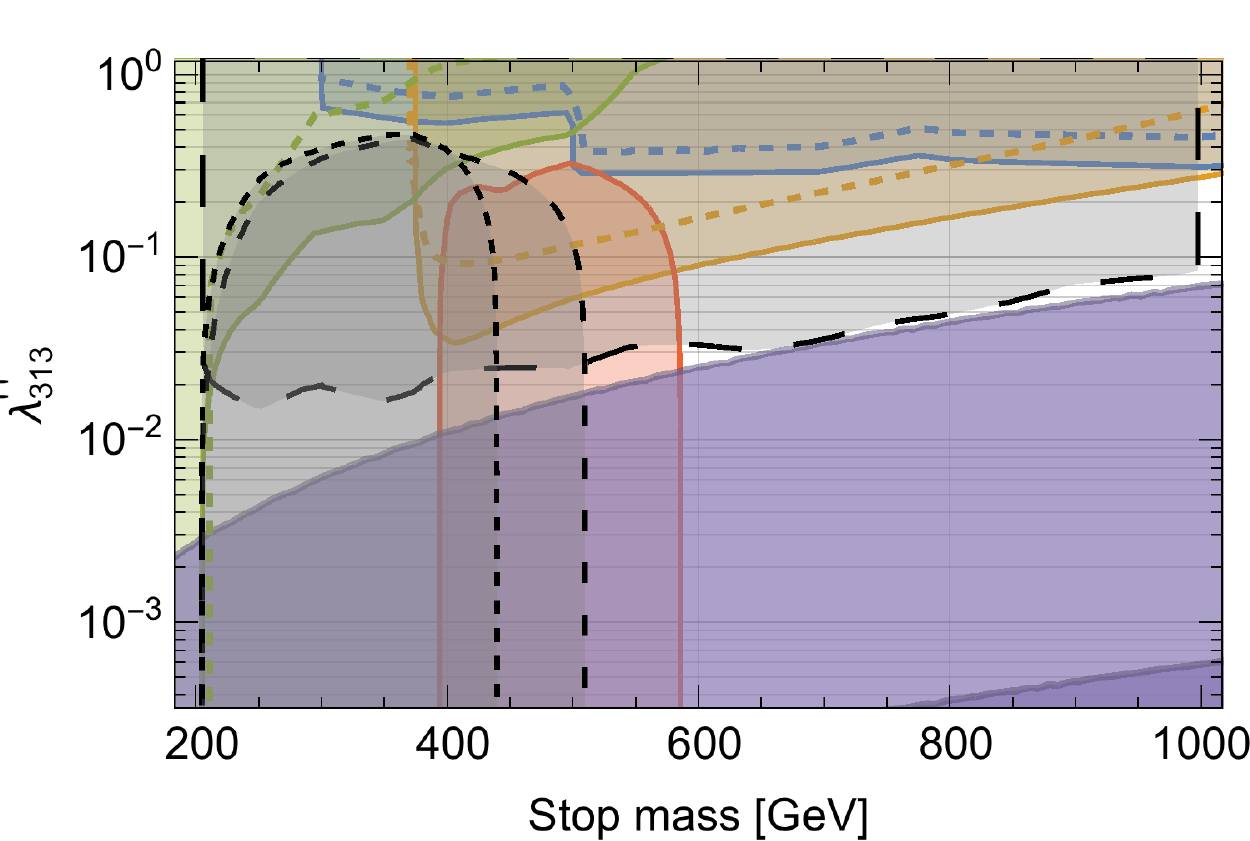}\\
\includegraphics[width=0.47\textwidth]{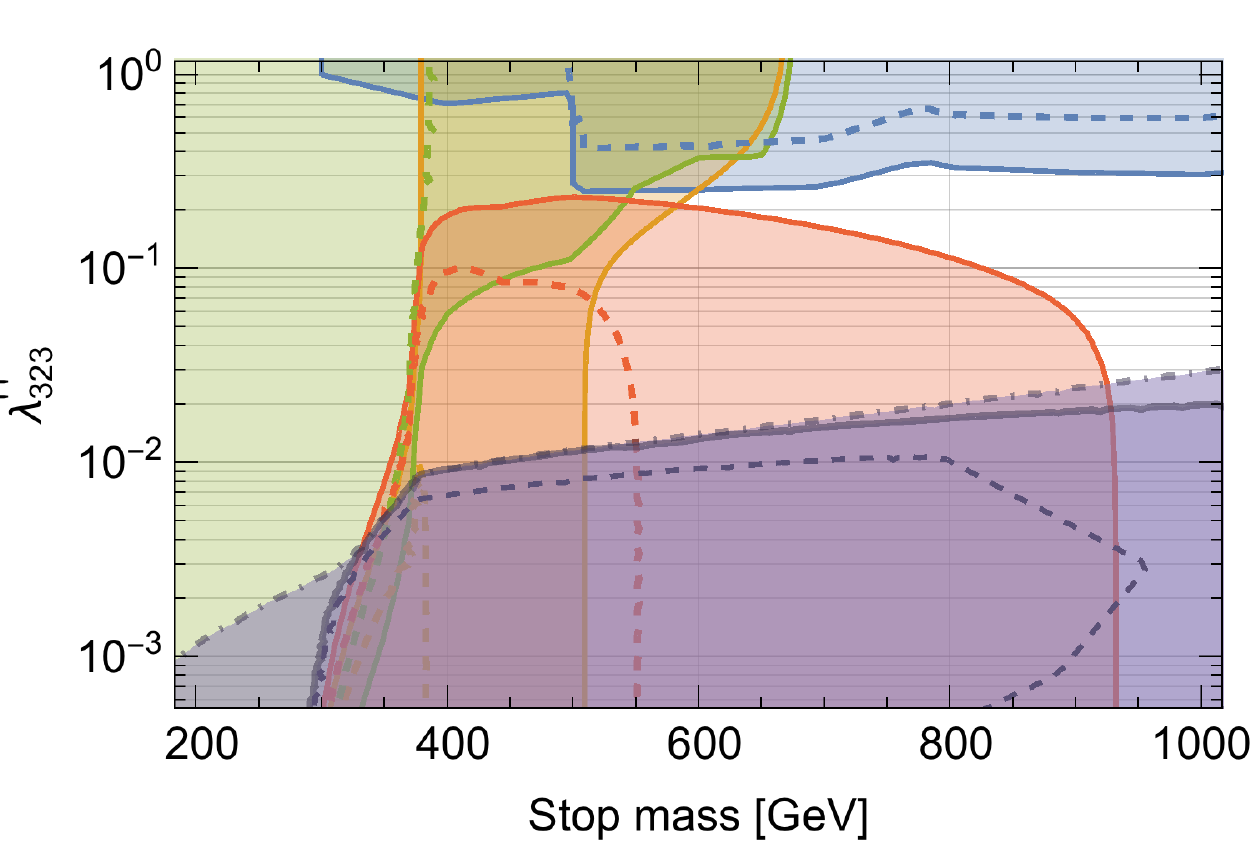}\qquad
\includegraphics[width=0.47\textwidth]{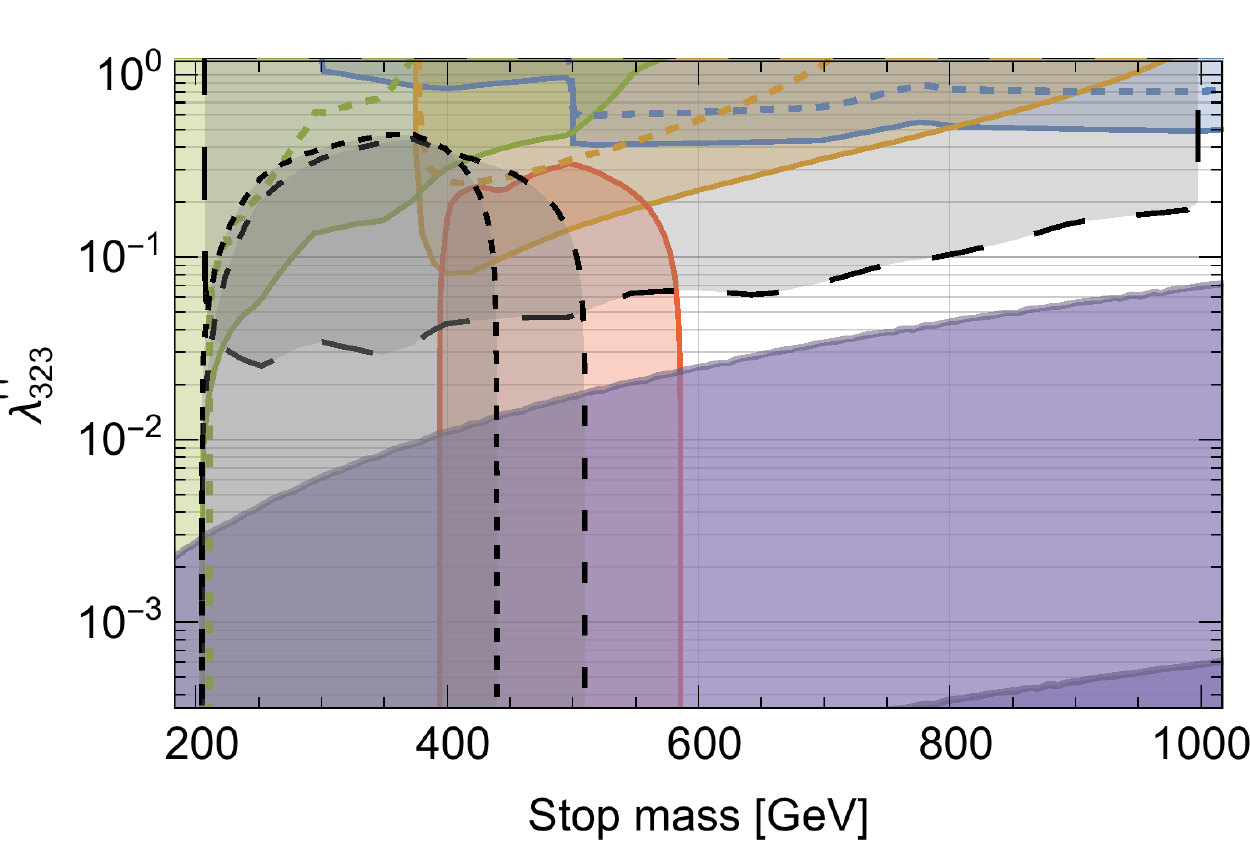}
\caption{Projected reach of LHC13 for each RPV coupling for a bino-like (left column) and for a higgsino-like neutralino LSP with $\mtn=200\gev$. Regions within dashed lines are already excluded by 8\tev\ data, while the reach of future searches (yielding hypothetical limits shown in the table at the top) is color-filled with matching color-coding.
}
\label{fig:lhc13}
\end{center}
\end{figure}}
\afterpage{\clearpage}

The following conclusions can be drawn:
\begin{itemize}
\item for a bino-like neutralino, existing searches (particularly same-sign top and four-top production) cover well most of the parameter space.  In particular, stop limits can be extended up to $\mtt\sim 900\gev$. Performing the same searches with the added requirement of two extra jets for each stop would likely improve the reach even more;
\item for a higgsino-like LSP, existing searches do not improve by much and most of the parameter space can only be probed via the chargino decay channel. In particular, pair-produced stops decaying to four quarks each could be excluded up to 500\gev, while the single-produced stop decaying in the same way can exclude $\ljk\gtrsim 0.1$ over the whole  sub-TeV stop mass range. Excluding very light charginos with prompt decays (three-jet resonances) would at the same time forbid long-lived neutral higgsinos.
\end{itemize}

\subsection{A generic SUSY spectrum}
So far, we have only discussed simplified scenarios in which all of the other superpartners are decoupled apart from the neutralino and the right-handed stop. While this is justified for the gluino, which is already excluded below 1\tev, one might worry that adding other superpartners below the stop mass might change our results; but, as long as a generic spectrum does not appreciably changes the right-handed stop or neutralino branching ratios, the limits discussed above would still hold. Furthermore, additional light superpartners could be probed via a wide range of signatures, which we now discuss.

For what concerns first- and second-generation squarks, the pair-production cross section is higher by a factor of order 10 and their dijets decays will be constrained by the paired dijet searches if the corresponding RPV couplings are not too small \cite{Khachatryan:2014lpa,ATLASCONF2015026}; in this channel, we find that the upper limit on the squark mass is already $m_{\tq_{1,2}}\gtrsim700\gev$. Otherwise, the \rpcing decay channel $\tilde q\to q\nino {}^{(*)}$, with the (possibly off-shell) bino-like neutralino decaying to $t d_j d_k$ via RPV would be important, and same-sign top signature would apply. This was studied in Ref. \cite{Durieux:2013uqa} where it was found that $m_{\tq_{1,2}}\gtrsim500\gev$. In the case of a higgsino-like LSP the squark decay to chargino, followed by $\chinop\to jjj$ would dominate and this has not been explicitly studied yet: we find that the CMS gluino search \cite{Chatrchyan:2013gia} excludes pair-produced squarks below 550\gev\ (with no $b$-jets in the chargino decay) and 600\gev\ (with one $b$-jet), if the squark branching ratio to charginos is nearly one. First and second-generation squarks would not be resonantly produced through the RPV couplings $\ljk$, so other large RPV couplings would need to be present for this channel to be relevant. While couplings involving multiple first-generation quarks are constrained by flavor physics, one could presumably have large couplings involving the charm squark, e.g. $\l_{223}$ (for example, in models with flavor symmetries one expects $\l_{223}\simeq \l_{312}\simeq 10^{-1}\l_{323}$).

A relatively light right-handed sbottom could enter the neutralino decay chain, but the final state would stay the same as the stop-mediated decay, $\nino\to b\tilde b^{*}\to b t d_k$. More importantly, another stop decay channel would open, $\st\to \tilde b W^+$, with the sbottom decaying via RPV $\tilde b \to \bar t \bar d_k$. Depending on the SUSY spectrum, this could suppress the same-sign top signature of RPV stops, but would also give multiple leptons (although opposite-sign and with a reduced number of jets). Resonant production of sbottom could not go through the $\ljk$ couplings due to the absence of $t$ in the proton PDF, but its production via the $\l_{2j3} C^c D_j^c B^c $ operator would have a yield similar to the stop production via $b s$ fusion (with the $\l_{323}$ coupling), modulo the different RPV couplings.

With \rpv, the LSP needs not to be electrically neutral and can also be a charged slepton. New  \rpcing decays appear, such as $\nino\to\ell\tilde \ell^*$ or $\chinop\to \nu_\ell\tilde \ell^*$ followed by $\tilde\ell^*\to\bar\ell t jj$. This usually gives same-sign dileptons, which could bring back constraints. An important possibility is a long-lived LSP slepton  with displaced vertices, as well as fake muon-like charged tracks in the detector \cite{Desai:2010sq}. In this case the SS top limits discussed earlier would disappear.

For a well-mixed neutralino sector, one would generally get a mixture of the higgsino- and bino-like scenarios discussed in this paper. As an example, taking a neutralino at 200\gev\ and a chargino at 400\gev, the bino-like limits discussed above apply until 400\gev\ after which the decay to a chargino dominates, giving multi-jet signatures for a stop decay. For a {\it natural} spectrum, one expects higgsinos and light charginos at $\mu\lesssim 250\gev$, meaning that stop decays via the chargino (with signatures outlined in Sec.~\ref{sec:new}) are theoretically well-motivated and should be investigated. If the neutralino and chargino masses are well-separated, the \rpcing chargino decay $\chinop \to W^+\nino$ will start to dominate at small $\ljk$: in this case the stop decay will still yield same-sign lepton signatures, $\st\to b\chinop\to b W^+\nino\to b W^+ (b W^+ ds )$, although with different kinematics than the same-sign top signature discussed in this work. In general, the wino component of the neutralino will only be relevant as long as the stop is not 100\% right-handed, otherwise it would be decoupled. The strength of the experimental signatures scales down as expected, with $\sigma(pp\to \st_1)\propto \cos\theta_\st^2$ and $\Gamma(\st_1\to f \tilde w)\propto \sin\theta_\st^2 M_{ij}^2$, where $M=N,U,V$ stands for either the neutralino or chargino mixing matrices.

\section{Conclusions}\label{sec:conclusions}
In this work, we have given multiple new constraints on large \rpving couplings involving the stop $\st$. We have shown that resonant squark production can be the dominant production mechanism and that limits on O(1) couplings apply well into the multi-TeV range. For what concerns the sub-TeV range, we have focused on same-sign top signatures arising from cascade stop decay into neutralinos and shown that they can probe RPV couplings well below $\ljk=0.1$. With a bino-like neutralino LSP,  NLSP stops below 500\gev\ are excluded apart from small slivers of parameter space. In the case of a higgsino-like LSP, stops lighter than 200\gev\ are still allowed and  searches targeting the stop decaying to a chargino are suggested in Section \ref{sec:new} (those also apply to a more generic neutralino sector and to first- and second-generation squarks). We have also shown that displaced neutralino decays are excluded for neutralino masses up to approximately 700\gev\ even in the bino case, where the small cross section for direct bino production is balanced by the large bino yield arising from stop decays. This limits on large \rpv\ can be used to exclude scenarios of baryogenesis as well as two-loops contributions of RPV stops to the Higgs mass \cite{Dreiner:2014lqa}.

With LHC Run 2 data already constraining New Physics in the multi-TeV region, the low-mass region should not be deemed as comprehensively covered. Long gone are the days when \rpv\ was a way to ``hide supersymmetry''. In fact, the reach from existent and proposed searches for resonantly produced squarks easily extends above the 1 TeV range, well above standard \rpcing SUSY searches. Given the novel signatures of the scenarios discussed, a discovery is still possible even at low masses.

\begin{acknowledgments}
The author would like to thank many colleagues for useful discussions and feedback during various stages of this work, in  particular David Shih, Scott Thomas, Matt Buckley, Eva Halkiadakis, Chang Sub Shin, Marco Farina, Simon Knapen. We also thank Tim Stefaniak for feedback regarding inconsistencies in Ref. \cite{Dreiner:2012np}. A.M. is  supported by the DOE grant DOE-SC0010008.
\end{acknowledgments}

\appendix
\section{Squark and neutralino decay rates}\label{sec:decayrates}
In this appendix we write down analytical expressions for decay rates used in the paper.

\paragraph{Stop}  First, we show the decay rates for the stop decay modes considered in this paper: for two-body decays where the final states are on-shell, we have
\beq
\Gamma_{\st\to d_j d_k}&=\frac{m_\st}{8\pi}|\l_{3jk}|^2\cos^2 \theta_\st\,,
\\
\Gamma_{\st\to t\nino}&=\frac \mtt{16\pi}\left[ (c_1^2+c_2^2) (1-x_t^2-x_{\nino}^2) -  4c_1c_2x_t|x_{\nino}|\right]\lambda^{1/2}(1,x_t^2,x_{\nino}^2)\,,\\
\Gamma_{\st\to b\chinop}&=\frac \mtt{16\pi } \left[(k_1^2+k_2^2) (1-x_b^2-x_{\chinop}^2) - 4 k_1k_2 x_b |x_{\chinop}| \right]\lambda^{1/2}(1,x_b^2,x_{\chinop}^2)\,,
\eeq
where $x_i\equiv m_i/\mtt$, $\lambda(a,b,c)\equiv a^2+b^2+c^2-2(ab+bc+ac)$ is the K\"all\'en function and the coefficients $c_i,k_i$ are the neutralino and chargino couplings to the stop (e.g. see \cite{Dreiner:2008tw}), which are functions of the squark mixing angles and the neutralino and chargino mixing matrices $N_{ij}$ and $U_{ij}, V_{ij}$:
\beq
c_1&= Y_t N_{14}\cos\theta_\st+\frac1{\sqrt2}(g N_{12}+\tfrac13 g' N_{11})\sin\theta_\st\,, \qquad c_2= Y_t N_{14}\sin\theta_\st-\frac{2\sqrt2}3g' N_{11}\cos\theta_\st\,,
\nn\\
k_1&= Y_t V_{i2}^*\cos\theta_\st - g V^*_{i1}\sin\theta_\st, \qquad\qquad
k_2= Y_b U_{i2}^*\sin\theta_\st \,.
\eeq
The phase-space suppression of the decay $\st\to t\nino$ is evident as $x_t=m_t/\mtt\gtrsim 0.2$ for sub-TeV stops, while $x_b\ll1$ for the chargino decay mode gives no suppression, $\lambda^{1/2}(1,x_b^2,x_{\chinop}^2)=(1-x_{\chinop}^2)$.

Four-body stop decays (with either an off-shell top quark or an off-shell neutralino) will be much more suppressed; decays with an off-shell top are relevant only near $\mtt \simeq m_t+\mtn$ when the dijet decay rate is small, $\ljk<10^{-3}$. The case with an off-shell neutralino which decays via RPV is discussed for example in the Appendix of Ref. \cite{Durieux:2013uqa}, where it is found that $\Gamma_{\st\to q q q q}\lesssim 10^{-6} \mtt|\l |^2$. This will always be sub-dominant with respect to the dijet decay rate $\Gamma_{\st\to qq}\simeq 0.04\mtt|\l|^2$. 

\paragraph{Neutralino}  The neutralino decay rate determines the neutralino decay length and can potentially result in displaced vertices. For a stop LSP, the decay rate for $\nino\to t\st$ is
\beq
\Gamma_{\tilde \chi\to t \tt^*} &=\frac{N_C}{32\pi}\mtn \lambda^{1/2}(1,y_t^2,y_\st^2)\left[ (c_1^2+c_2^2) (1+y_t^2-y_\st^2)+ 4c_1c_2 y_t \right]\,,
\eeq
where $y_i\equiv m_i/\mtn$ and $c_1,c_2$ are defined above.
For a neutralino LSP the decay goes through an off-shell stop and the coupling $\ljk$:
\beq\label{eq:ninodecayrate}
\Gamma_{\nino\to{t d_j d_k}}=\frac{\mtn^5}{ 1024\pi^3\mtt^4}|\ljk|^2 \bigg((c_1^2+c_2^2)   I_1 (y_t,y_\st)+4 c_1c_2 y_t\, I_2(y_t,y_\st)\bigg)\,.
\eeq
Here $y_i\equiv m_i/\mtn$ and  $I_1, I_2$ are phase-space integrals defined as
\beq
\system{I_1(y_t,y_\st)\\I_2(y_t,y_\st)}\equiv \int_{2y_t}^{1+y_t^2} dz_t\frac{(1-z_t+y_t^2)(z_t^2-4y_t^2)^{1/2}}{(1-(1-z_t+y_t^2)/y_\st^2)^2}\system{12 z_t\\\ 6}\,,
\eeq
which satisfy $\lim_{\substack{y_t\to0\\ y_\st\to\infty}} I_{1,2}(y_t,y_\st)=1$. They can be expanded analytically in the massless top quark limit  ($y_t\ll1$) or in the large stop mass limit ($y_\st=1/{x_\nino}\gg1$):
\beq
&\lim_{y_\st\to\infty}I_1(y,y_\st)\equiv f(y)=1-8y^2+8y^6-y^8-24y^4\log y,\\
&\lim_{y_\st\to\infty}I_2(y,y_\st)\equiv g(y)=1+9y^2-9y^4-y^6+12(y^2+y^4)\log y,
\\\nn
&\lim_{y_t\to0}I_1(y_t,\tfrac1{x^2}) \equiv h(x)= 6\frac{6 x^2-5 x^4+2 \left(3-4 x^2+x^4\right) \log \left(1-x^2\right)}{x^8}\simeq1+\frac45x^2+\frac35x^4+\ldots
\eeq
For $y_t\to0$, $I_2$ takes a similar expression to $I_1$, but this is not relevant as it appears in the decay rate as $y_t\cdot I_2\to0$.
The difference between the full integral $I_1$ and the product of the approximate expressions, $f\cdot h$, is small in most of the parameter space.

If the neutralino is lighter than the top quark, the decay rate will be further suppressed by the off-shell top and the four-body phase space, $\nino\to b W^+ d_j d_k$. In this case we compute the decay rate numerically with \texttt{MadGraph5}. One can see in Fig. \ref{fig:ninoctau} the parameter region where the neutralino give prompt decays, displaced decays or is long-lived: appreciable decay lengths usually happen for $\mtn\gtrsim m_t$.

\paragraph{Chargino} The decay rate for the chargino, $\chinop\to \bar b\bar  d_j\bar  d_k$, takes a similar form as for the neutralino, Eq. \eqref{eq:ninodecayrate}, except for the exchange $c_i\to k_i$ and the absence of the top quark phase-space suppression, $y_b\simeq0$ and $I_1(y_b,y_\st)\simeq h(\mtch/\mtt)$: 
\beq
\Gamma_{\chinop\to \bar b\bar  d_j\bar  d_k}=\frac{\mtch^5}{ 1024\pi^3\mtt^4}|\ljk|^2 (k_1^2+k_2^2) h(\mtch/\mtt)\,.
\eeq
Again, one can see in Fig. \ref{fig:ninoctau} the regions where displaced chargino decays are possible.

\section{Dijet acceptances}\label{sec:accept}
In this appendix, we describe in details the procedure to find the signal acceptances for the ATLAS and CMS dijet searches used in this work.

For the ATLAS searches \cite{Aad:2014aqa,ATLAS:2015nsi}, we follow the procedure outlined in the Appendix of \cite{Aad:2014aqa} to set limits on our model: first, for a given stop mass $M$ and for each coupling $\ljk$ we generate 20,000 events with \texttt{MadGraph5} for the resonant production and subsequent decay, $pp\to \st\to jj$,  after which we apply the kinematic cuts used in the analysis:
\beq
&\text{ATLAS \cite{Aad:2014aqa} ($\sqrt s=8\tev$)}: \quad |\eta|<2.8,\quad p_T> 50\gev,\quad m_{jj}>250\gev,\quad |\Delta\eta_{jj}|<1.2\nn\,,\\
&\text{ATLAS \cite{ATLAS:2015nsi} ($\sqrt s=13\tev$)}: \quad |\eta|<2.8,\quad p_T> 440\gev,\quad m_{jj}>1.1\tev,\quad |\Delta\eta_{jj}|<1.2\,.\qquad
\eeq
As the search is for a narrow Gaussian resonance, we remove from the generated signal  the tails away from the mass $M$ of the resonance, only keeping events with invariant dijet mass in the range $0.8M<m_{jj}<1.2M$. The fraction of events left in the sample defines the modified acceptance $A$, which is tabulated in Table \ref{tab:accept},  and goes from about $0.5$ for $M=300\gev$ to $0.02$ for $M=4\tev$ (at $\sqrt s8\tev$). To find the limit on the cross section for each value of $M$, we perform a Gaussian fit to the truncated dijet invariant mass distribution, which yields the reconstructed mass $m_G$ and width $\sigma_G$ of the dijet resonance; the width-to-mass ratio is about $5-7\%$ over the range of stop masses, comparable to the dijet mass resolution. Then, for a given RPV coupling we compare the ATLAS experimental  95\% C.L. upper limit on $\sigma\times A$ for the chosen mass $m_G$ and width $\sigma_G/m_G$ to the resonant production cross section times the modified acceptance obtained earlier. Values of $\ljk$ yielding cross sections above the limits are excluded.

For the CMS searches \cite{Khachatryan:2015sja,CMS-PAS-EXO-14-005,Khachatryan:2015dcf}, ``wide jets'' are constructed from geometrically close jets (jets within $\Delta R=\sqrt{(\Delta\eta)^2+(\Delta\phi)^2}<1.1$ of the highest $p_T$ jet are added to it iteratively) and the full dijet invariant mass distribution is kept, allowing to distinguish between quark-quark, quark-gluon and gluon-gluon resonances (gluon-seeded wide jets have larger tails). Results are given as 95\% C.L. upper limits on $\sigma\times Br_{qq} \times A$, where in this case the acceptance $A$ is the fraction of events surviving the kinematic cuts, and is larger than in the ATLAS searches described above. For all searches, shared kinematic cuts are $|\eta|<2.5,\, p_T>30\gev$, while the other cuts are:
\beq
&\text{CMS \cite{Khachatryan:2015sja} }( \sqrt s=8\tev):\quad H_T> 650\gev,\quad m_{jj}>890\gev,\quad |\Delta\eta_{jj}|<1.5\;;\\
&\text{CMS \cite{Khachatryan:2015dcf} } (\sqrt s=13\tev):\quad H_T> 800\gev,\quad m_{jj}>1.2\tev,\quad |\Delta\eta_{jj}|<1.3\;;\nn\\
&\text{CMS \cite{CMS-PAS-EXO-14-005} } (\text{scouting, } \sqrt s=8\tev): \quad H_T> 250\gev,\quad m_{jj}>390\gev,\quad |\Delta\eta_{jj}|<1.3.\nn
\eeq
These searches  produce limits on $\sigma\times Br_{qq}\times A$ in the ranges 1.2\tev-5.5\tev, 1.5\tev-7\tev\  and 500\gev-1.2\tev, respectively. The last one, \texttt{CMS-PAS-EXO-14-005} \cite{CMS-PAS-EXO-14-005} is able to give limits in the low mass region where the event rate is large (due to QCD background) by a using a {\it data scouting stream}, where instead of storing all the full event information, only four-vectors of reconstructed jets are saved. In this way, the required per-event storage  is smaller and a larger event rate can be achieved.

\begin{table}[t]
\begin{center}
\begin{tabular}{c|cccc}\hline\hline
& \multicolumn{4}{c}{Acceptance for each dijet analysis}
\\ 
 Stop mass [GeV] & ATLAS$_{8\tev}$ \cite{Aad:2014aqa} & CMS$_{8\tev}$ \cite{Khachatryan:2015sja,CMS-PAS-EXO-14-005} & ATLAS$_{13\tev}$ \cite{ATLAS:2015nsi} &CMS$_{13\tev}$ \cite{Khachatryan:2015dcf} \\\hline 
 300 & 0.50 & - & - & - \\
 400 & 0.50 & - & - & - \\
 500 & 0.48 & 0.52 & - & - \\
 600 & 0.46 & 0.52 & - & - \\
 700 & 0.45 & 0.52 & - & - \\
 800 & 0.43 & 0.52 & - & - \\
 900 & 0.42 & 0.52 & - & - \\
 1000 & 0.41 & 0.51 & - & - \\
 1200 & 0.38 & 0.48 & 0.38 & - \\
 1500 & 0.34 & 0.46 & 0.41 & 0.50 \\
 1800 & 0.30 & 0.43 & 0.38 & 0.51 \\
 2000 & 0.27 & 0.40 & 0.37 & 0.50 \\
 2400 & 0.19 & 0.35 & 0.33 & 0.49 \\
 2800 & 0.13 & 0.28 & 0.30 & 0.47 \\
 3200 & 0.070 & 0.23 & 0.26 & 0.44 \\
 3600 & 0.040 & 0.19 & 0.21 & 0.42 \\
 4000 & 0.020 & 0.16 & 0.17 & 0.38 \\\hline\hline
\end{tabular}
\caption{Modified acceptances for the ATLAS Gaussian dijet resonance searches \cite{Aad:2014aqa,ATLAS:2015nsi}, as described in the text, and acceptances for the CMS wide jet searches \cite{Khachatryan:2015sja,CMS-PAS-EXO-14-005,Khachatryan:2015dcf}, both at $\sqrt s=8\tev$ and $\sqrt s=13\tev$ .}
\label{tab:accept}
\end{center}
\end{table}%
\begin{figure}[t]
\begin{center}
\includegraphics{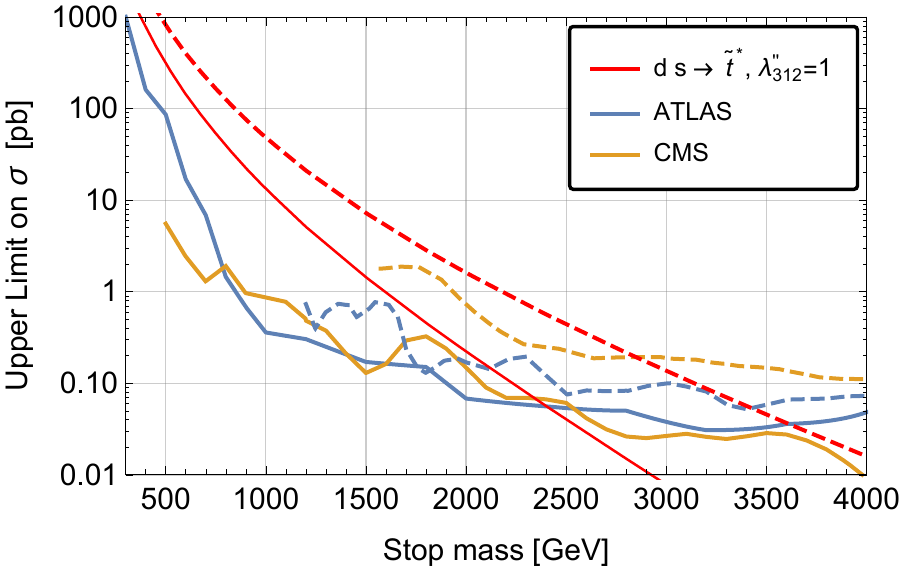}
\caption{Most important experimental limits on the production cross section of a dijet resonance. Limits from ATLAS narrow resonance searches \cite{Aad:2014aqa,ATLAS:2015nsi} are shown in blue while limits from CMS wide jet searches \cite{Khachatryan:2015sja,CMS-PAS-EXO-14-005,Khachatryan:2015dcf} are in orange; for reference, in red we show the cross section for the resonant production via the $\l_{312}$ coupling (set to unity), $ds\to \st^*, \ \bar d\bar s\to \st$. solid lines refer to collisions at $\sqrt s=8\tev$ while dashed lines refer to $\sqrt s=13\tev$.}
\label{fig:sigma_dijets}
\end{center}
\end{figure}

In Fig. \ref{fig:sigma_dijets}, we take the experimental limits on $\sigma \times A$ and, using the acceptances in Table \ref{tab:accept}, show the limits on the production cross section to easily identify the most constraining searches. In most of the mass range the ATLAS searches give better limits, with exceptions for the CMS {\it scouting} analysis \cite{CMS-PAS-EXO-14-005} below 1\tev. In red we show the production cross section via the $\l_{312}$ coupling. Preliminary results with 13\tev\ data are shown as dashed lines (note the increase in the resonant production cross section). The dijets limits on RPV presented in the main body of the paper are based on the best available limits at each mass.

\section{More on displaced vertices}\label{app:dv}

In this section, we discuss in more details the limits on displaced decays of neutralinos and charginos. We use results from Refs.~\cite{Cui:2014twa,Liu:2015bma,Csaki:2015uza} which recasted the original experimental searches \cite{Aad:2015rba,CMS:2014wda}.

In particular, our starting point is Section 4.1 of Ref.~\cite{Cui:2014twa}, which is based on the CMS displaced dijet analysis \cite{CMS:2014wda}. There, limits on the displaced decays of pair-produced winos (each decaying to three jets) are given. We reproduce the 95\% CL upper limits on the cross section (including the comments in Section 4.1.1) in Fig.~\ref{fig:crossx_dv} for different lifetimes in the range $1\text{ mm}<c\tau<10 \text{ m}$. As a cross check, we compare these limits to the higgsino cross section, find the excluded region in the $m_{\tilde H}-c\tau_{\tilde H}$ plane and show in Fig. \ref{fig:hinoctau_comparison} the comparison with the results of Ref.~\cite[Fig.~5]{Liu:2015bma} , where constraints on displaced higgsino decays were studied: the excluded regions are similar, particularly in the low mass region ($m_{\tilde H}\lesssim 300\gev$) where displaced decays can take place with appreciable RPV couplings $\ljk\gtrsim 10^{-4}$.\footnote{For $\mtn\lesssim m_t$ neutralino decays go through an off-shell top, and for $\mtn \lesssim 300\gev$ the top-quark phase space suppression is large. At higher masses there is little phase space suppression and displaced decays are only possible for small RPV couplings, $\ljk\ll10^{-6}$.
}
The small differences can be attributed to the different recasting procedures of Refs.~\cite{Cui:2014twa,Liu:2015bma}. In particular, using the results of Ref.~\cite{Cui:2014twa} gives slightly weaker constraints: as recasting the original experimental searches is beyond the scope of this work, we will be conservative and use those limits. 

\begin{figure}[t]
\begin{center}
\subfloat[]{\includegraphics[width=0.45\textwidth]{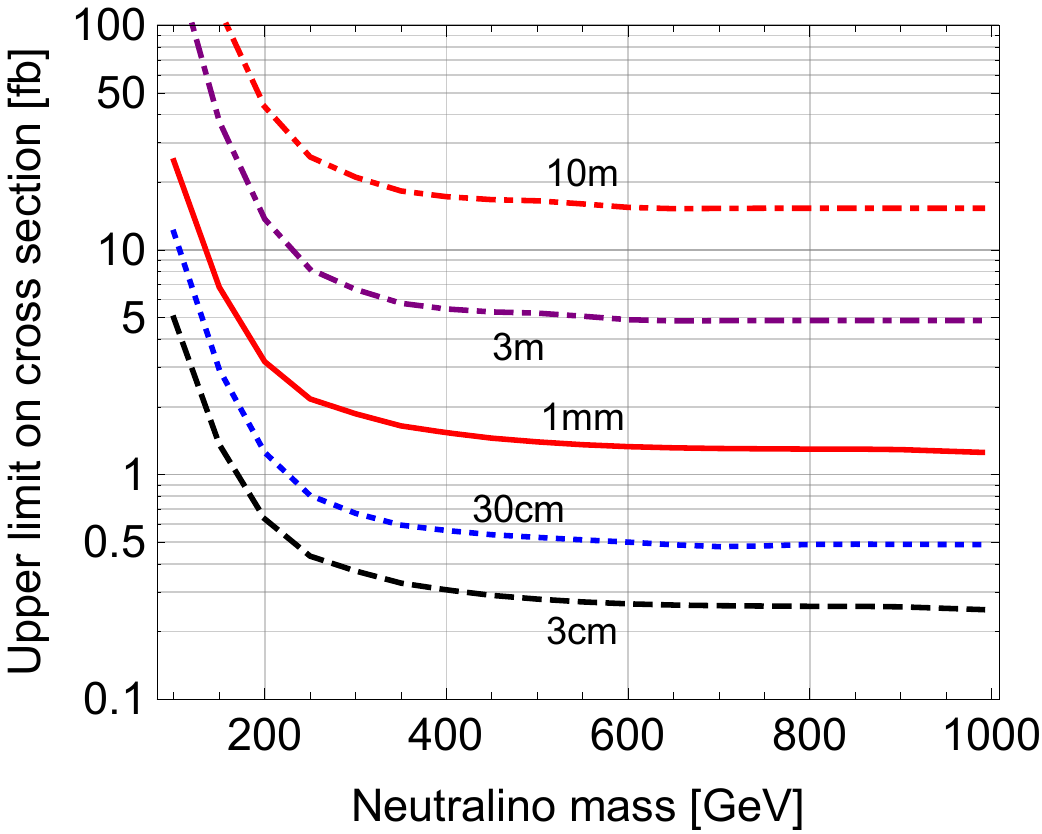}\label{fig:crossx_dv}}
\qquad \subfloat[]{\includegraphics[width=0.45\textwidth]{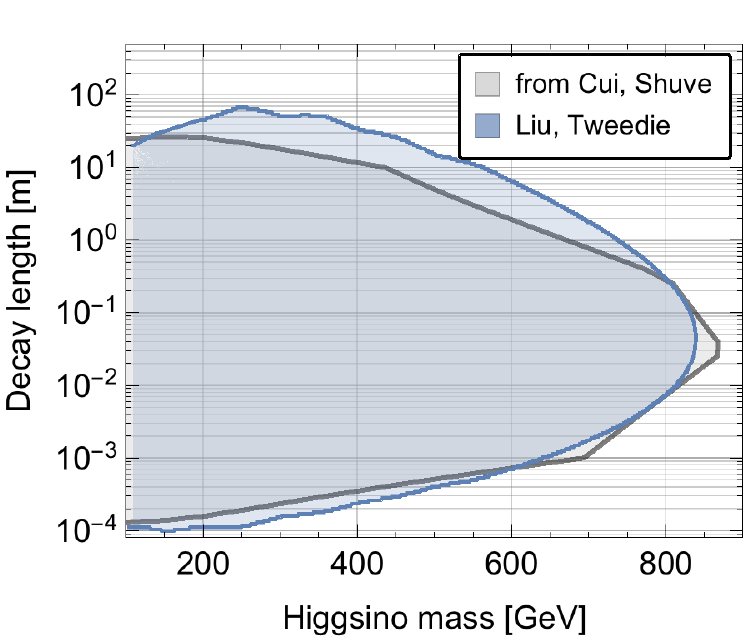}\label{fig:hinoctau_comparison}}
\caption{Left: Upper limits on the pair-production cross section of a three-jet resonance (in our case, the neutralino) as a function of its mass and of its lifetime, based on Ref.~\cite{Cui:2014twa}. Right: Comparison between our results and those of Ref.~\cite{Liu:2015bma} for displaced higgsinos: the region enclosed by the gray (blue) line is excluded by our (Ref. \cite{Liu:2015bma}'s) analysis.}
\end{center}
\end{figure}

The original studies that we are discussing were focused on three-jet decays, $X\to j j j$, while in this work we are mostly interested in decays involving the top, $X\to tjj$, followed by the instantaneous $t\to b W$ and $W\to jj, \ell \nu$ decays: then, one recovers a three-jet final state, with the possibility of five jets when the $W$ decays hadronically (if the decays are displaced enough, $c\tau\gtrsim 0.3$ mm, the $b$ decays can be considered instantaneous and will give either a $b$-jet or a jet+muon, see also \cite[Sec. 3.1]{Liu:2015bma}). Thus, if anything we expect the limits on decays to top quarks to be stronger and we use the conservative limits leading to Fig.~\ref{fig:hinoctau_comparison} to set limits on displaced neutralino decays, $\nino\to tjj$.

Finally, in this work we have also studied displaced decays of neutralinos arising from stop decays: for example, the bino direct pair-production cross section is rather small, and also depends on first- and second- generation squark masses, which are probed by other multijet searches and we have chosen to decouple from the light spectrum. But even if direct pair-production is small, stop pair-production followed by the decay $\st\to t\nino$ can produce a large amount of binos if the channel is kinematically accessible. We are then able to exclude regions of the parameters space where displaced binos originate from stops. In addition, if the bino is stable on collider timescales  it exits the detector and the usual \rpcing SUSY searches apply. For a higgsino, there are two differences: the direct pair-production cross section of both charged and neutral higgsinos is sizable, $\sigma\gtrsim 1$ fb, and the stop decay $\st\to b\chinop$ is preferred to $\st\to t\nino$. Then, the direct pair-production is enough to exclude both neutralino and chargino displaced decays (in addition, collider-stable charginos are excluded by HSCP searches). In general the neutralino lifetime is always larger than the chargino lifetime, especially for $\mtn\lesssim m_t$. We show the regions excluded by neutralino and chargino decays in Fig. \ref{fig:dv_hino}, for different choices of the bino and higgsino masses in the range $100\gev\leq \mtn\leq 200\gev$. For the higgsino, it can be seen that displaced chargino decays  also give a {\it lower limit } on the size of the RPV coupling, which is in the range $10^{-5}-10^{-2}$ for higgsinos below 300\gev.

\begin{figure}[tbhp]
\begin{center}
\includegraphics[width=\textwidth]{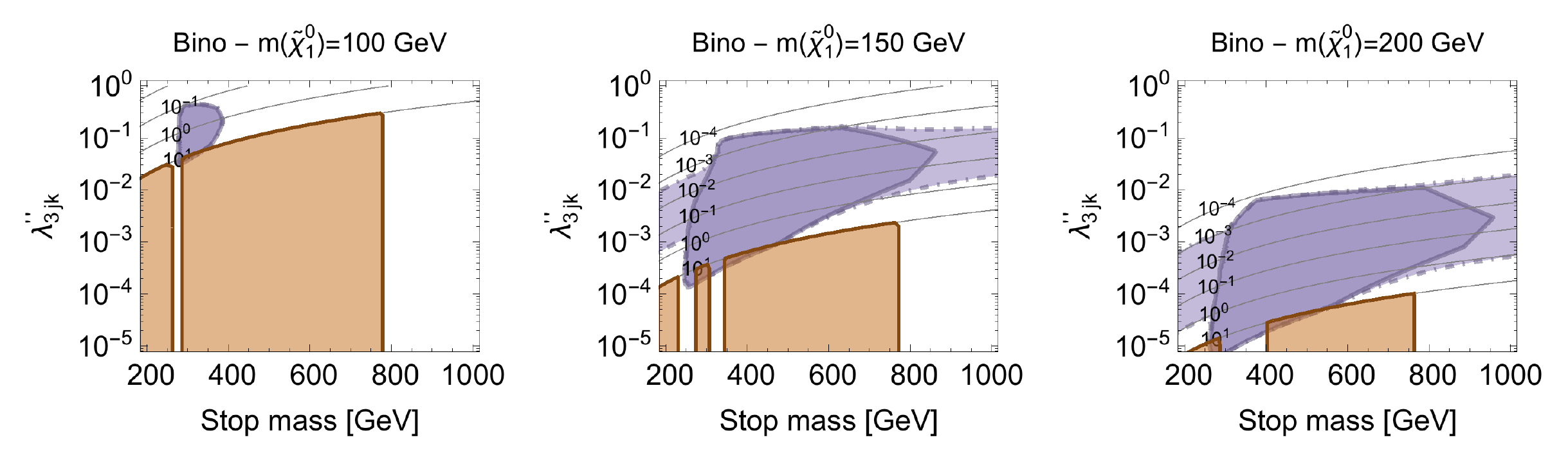}
\includegraphics[width=\textwidth]{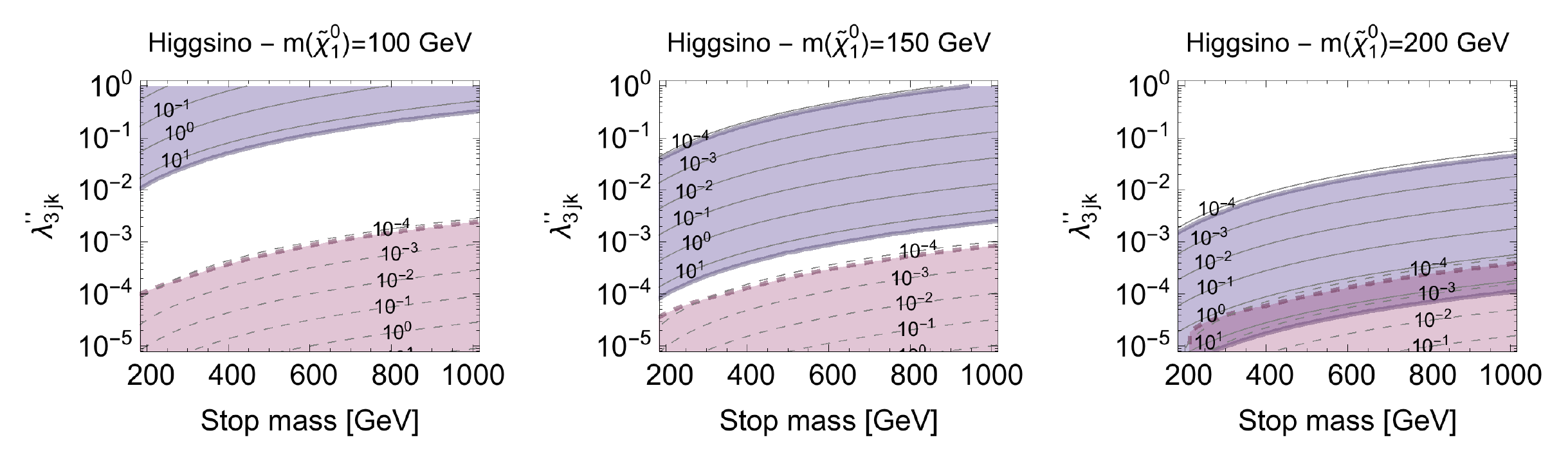}
\caption{Top: regions excluded by displaced and long-lived binos; dark blue regions enclosed by solid (dot-dashed) lines have $m_{\tq_{1,2}}=1 (10) \tev$, while orange regions at the bottom are excluded by RPC searches.
Bottom: regions excluded by displaced higgsino decays; the purple region corresponds to a displaced neutralino, while the magenta region below the dashed lines corresponds to a displaced or collider-stable chargino.}
\label{fig:dv_hino}
\end{center}
\end{figure}


\providecommand{\href}[2]{#2}\begingroup\raggedright\endgroup

\end{document}